\documentclass[aps,amsmath,amssymb, superscriptaddress, twocolumn, showpacs, longbibliography]{revtex4-1}
\usepackage[english]{babel}
\usepackage{graphicx}
\usepackage{amsmath}
\usepackage{amssymb}
\usepackage[outdir=./]{epstopdf}
\usepackage{amsfonts}
\usepackage{bm}
\usepackage{mathtools}
\usepackage{times}
\usepackage{hyperref}
\usepackage{color}
\newcommand{\ket}[1]{|#1\rangle}
\newcommand{\bra}[1]{\langle#1|}

\newcommand{\beq}[1]{\begin{equation} #1 \end{equation}}
\newcommand{\bsplit}[1]{\begin{equation} \begin{split} #1 \end{split} \end{equation}}

\newcommand{\astcycl}{\mathrlap{\kern0.085em{\circlearrowright}}\ast}
\newcommand{\taucycl}{\mathrlap{\kern0.42em{\bullet}}\circlearrowright}
\long\def\/*#1*/{}

\newcommand{\+}{\dagger}
\def\<{\left\langle}
\def\>{\right\rangle}
\def\up{\uparrow}
\def\dn{\downarrow}

\def\P2d2h{\mathrm{P}_\mathrm{2d2h}}
\def\Pso{\mathrm{P}_\mathrm{so}}
\def\Pbexc{\mathrm{P}_\text{bexc}}
\def\Pexc{\mathrm{P}_\text{exc}}
\def\Phf{\mathrm{P}_\text{hf}}
\def\hD{\hat{D}}
\def\hh{\hat{h}}
\def\hc{\hat{c}}
\def\hn{\hat{n}}

\newcommand{\vk}{\textbf{k}} \newcommand{\vK}{\textbf{K}}

\newcommand{\vq}{\textbf{q}} \newcommand{\vQ}{\textbf{Q}}
\newcommand{\vR}{\textbf{R}}

\newcommand{\vx}{\textbf{x}} \newcommand{\vX}{\textbf{X}}

\setcitestyle{numbers,square}													

\begin{document}

\title{Photo-enhanced excitonic correlations in a Mott insulator with nonlocal interactions}
\author{Nikolaj Bittner}
\email{nikolaj.bittner@unifr.ch}
\affiliation{Department of Physics, University of Fribourg, 1700 Fribourg, Switzerland}
\author{Denis Gole\v{z}}
\affiliation{Department of Physics, University of Fribourg, 1700 Fribourg, Switzerland}
\affiliation{Flatiron Institute, Simons Foundation, 162 Fifth Avenue, New York, NY 10010, USA}
\author{Martin Eckstein}
\affiliation{Department of Physics, University of Erlangen-N\"urnberg, 91058 Erlangen, Germany}
\author{Philipp Werner}
\email{philipp.werner@unifr.ch}
\affiliation{Department of Physics, University of Fribourg, 1700 Fribourg, Switzerland}

\begin{abstract}
We investigate the effect of nonlocal interactions on  the photo-doped Mott insulating state of the two-dimensional Hubbard model using a nonequilibrium generalization of the dynamical cluster approximation.  In particular, we compare the situation where the excitonic states are lying within the continuum of doublon-holon excitations to a set-up where the excitons appear within the Mott gap. In the first case, the creation of nearest-neighbor doublon-holon pairs by excitations across the Mott gap results in enhanced excitonic correlations, but these excitons quickly decay into uncorrelated doublons and holons. In the second case, photo-excitation results in long-lived excitonic states. While in a low-temperature equilibrium state, excitonic features are usually not evident in single-particle observables such as the photoemission spectrum, we show that the photo-excited nonequilibrium system can exhibit in-gap states associated with the excitons.
 The comparison with exact-diagonalization results for small clusters allows us to identify the signatures of the excitons in the photo-emission spectrum.
\end{abstract}


\maketitle
\section{Introduction}

The photo-excitation of electrons across a Mott gap generates mobile charge carriers and thereby turns a correlation induced insulator into a nonthermal metal. This simple example of a light-induced phase transition has been studied experimentally for many years. Time-resolved measurements of the
optical response~\cite{iwai2003,okamoto2010,novelli2014}
and photo-emission spectrum~\cite{perfetti2006,ligges2018,rameau2016} have revealed the timescales associated with several physical processes, including the formation of the metallic state, gap renormalization, intra-band relaxation, polaron formation, and carrier recombination. On the theory side, many fundamental properties of photo-doped insulators have been investigated and clarified in studies of simple model systems, such as the single-band Hubbard\cite{eckstein2011,eckstein2016}, Holstein-Hubbard~\cite{werner2015} and \mbox{t-J} model~\cite{zala2013,bittner2018}. This includes the effect of electron-phonon and electron-spin interactions on the energy dissipation of photo-doped carriers~\cite{golez2014,eckstein2016}, impact ionization\cite{werner2014,sorantin2018}, and the gap size dependence of the recombination time~\cite{eckstein2011}.

As the theoretical effort shifts from the study of simple models to more realistic descriptions of photo-excited materials, one aspect which needs to be considered is the long-ranged nature of the Coulomb interaction. In a photo-excited single-band Mott insulator, the charge carriers are doublons (doubly occupied sites) and holons (empty sites) that move in a half-filled Mott background. In the presence of a nonlocal interaction, these doublons and holons can form bound states (excitons) which may affect the nature of the photo-doped state and the relaxation dynamics. Photoinduced excitonic features have been studied using exact diagonalization of small lattice systems~\cite{lu2015,shinjo2017}. These excitonic features can originate either from the non-local interactions~\cite{shinjo2017,zala2015, mitrano2014}
or the modification of the spin background~\cite{zala2013}.  Methods for infinite lattices based on single-site dynamical mean field theory (DMFT)~\cite{georges1996}, such as extended DMFT~\cite{sun2002, ayral2013, golez2015, werner2016b} or the combination of GW and extended DMFT~\cite{biermann2003, werner2016a, golez2017}, can capture the dynamical screening of the local Coulomb interaction resulting from nonlocal interactions, but they cannot describe exciton formation.
Here, we combine the two approaches by implementing a cluster extension of nonequilibrium DMFT~\cite{eckstein2016,tsuji2014}, with local and nonlocal interactions on the cluster.
Specifically, we consider the $U$-$V$ Hubbard model on the square lattice and the so-called dynamical cluster approximation (DCA)~\cite{hettler1998, maier2005} with a periodized cluster of four sites. For this model, a recent equilibrium study has demonstrated a fast convergence of the results with cluster size~\cite{terletska2017}.  The nonequilibrium DCA approach allows us to measure electron-hole correlations on the periodized four-site cluster in photo-doped states, and to connect these results with other observables such as the photoemission spectrum.
\begin{figure*}[ht]
\centering
\includegraphics[width=0.8\textwidth]{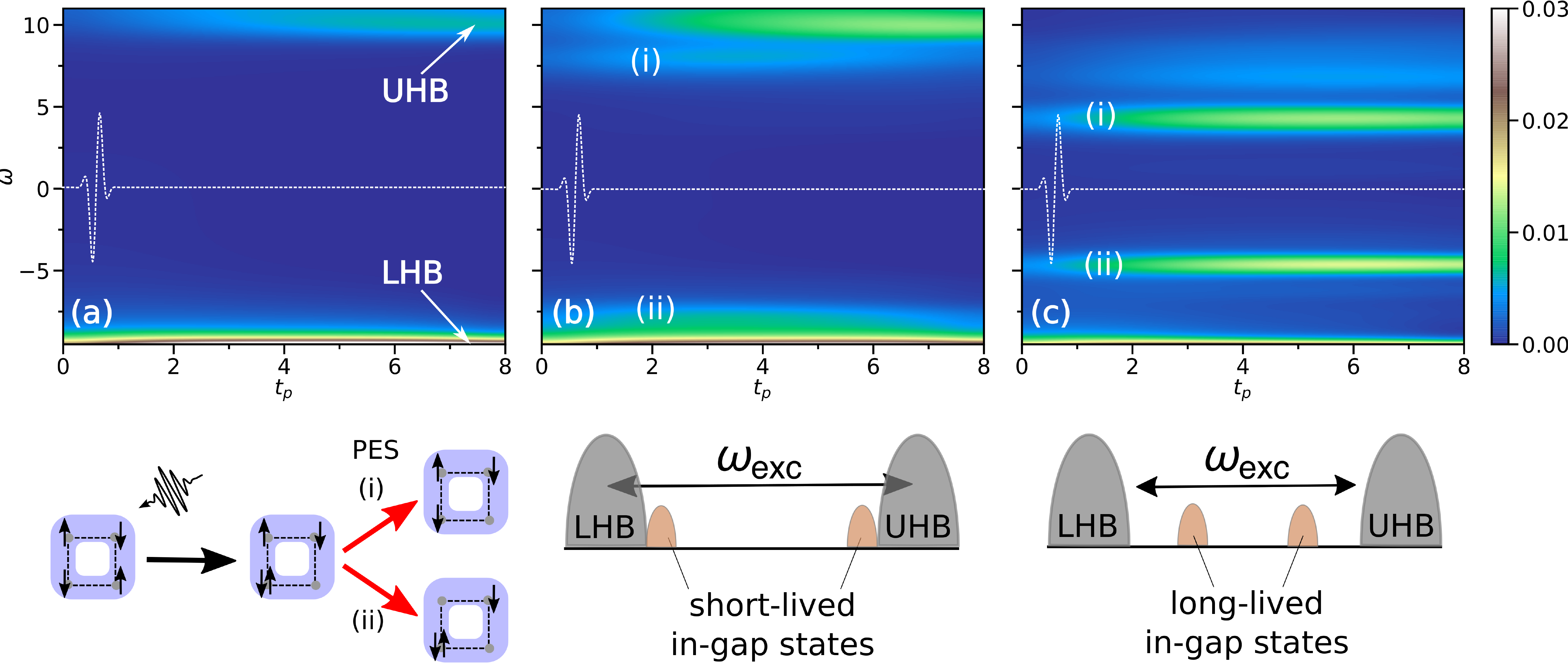}
\caption{(color online) Time-dependent spectral function
[$I(\omega)$ defined in Eq.~(\ref{eq:spectrum})] for $U=25$, and (a) $V=0$, (b) $V=3$, (c) $V=6$ within the Mott-gap region after a single-cycle photo-excitation with frequency $\omega=20$ (white dashed line). The lower (upper) Hubbard bands are indicated by LHB (UHB).
The peaks labelled by (i) and (ii) indicate photoemission processes from the photo-excited (nearest neighbor) exciton as indicated by the sketch. 
While for $V=0$ only the UHB is partially populated after the pulse, for $V>0$ in-gap states
appear immediately after the photo-excitation. Rough real-space sketches of dominant contributions with a nearest-neighbor exciton~(second row  left) illustrate the photo-emission process~(red arrow) corresponding to the signal (i) [removal of an electron from the exciton] and (ii) [removal of an electron which leaves the exciton intact]. The blue background in the  sketches represents the corresponding wave functions. 
For $V=3$ the excitonic states couple to the continuum of doublon-holon excitations which results in a fast decay. For $V=6$ the excitonic states are isolated within the Mott gap, resulting in long-lived excitonic features in the spectral function.  For all cases the duration of the probing pulse is set to $\Delta t_\mathrm{probe}=3$.
}
\label{fig:Ipes3D}
\end{figure*}

Our main finding is that the nearest-neighbor interaction $V$ leads to enhanced excitonic correlations compared to a chemically doped state. For sufficiently large Mott gap and nearest-neighbor interaction $V$, the nonequilibrium population results in prominent in-gap peaks in the
time-dependent spectral function, as summarized in Fig.~\ref{fig:Ipes3D}.
We have identified two characteristic non-equilibrium regimes: short-lived excitons are formed for weak to intermediate nearest-neighbor interaction $V$, while for strong $V$ the photo-induced excitons are long-lived. The transition  between the two regimes happens when the nearest-neighbor interaction $V$ shifts the exciton energy out of the doublon-holon continuum, compare Fig.~\ref{fig:Ipes3D}(b) and (c), which demonstrates the crucial effect of non-local interactions on the life-time and nature of photo-induced states. The analysis of the real-space correlation functions reveals that excitons are formed dominantly on nearest-neighbor sites. While next-nearest neighbor excitons and biexcitons are also enhanced by an external pulse, their density is at least an order of magnitude smaller than for the nearest-neighbor excitons.

The paper is organized as follows. Section~\ref{sec:Model} describes the model and observables.  Sec.~\ref{sec:Results} presents an analysis of various correlation functions and the local spectral function both in equilibrium and in a photo-doped state. Section~\ref{sec:Summary} summarizes the main results. A detailed description of the DCA method used to solve the model is present in Appendix~\ref{sec:Method}.

\section{Model and observables}
\label{sec:Model}

\subsection{Model and method}
We simulate a half-filled strongly correlated electron system with on-site interaction $U$ and nearest-neighbor interaction $V$ on a two-dimensional (2D) square lattice. The Hamiltonian of this so-called extended Hubbard model (EHM) can be written as
\begin{align}
\label{eq:EHM}
	H(t)=&- \sum_{\<i,j\>\sigma}\left[ t_h (t) \hc_{i\sigma}^\+\hc_{j\sigma} + h.c. \right] - \mu \sum_{i} \hn_i \nonumber\\
	   & +U\sum_i \hn_{i\up}\hn_{i\dn} +V\sum_{\<i,j\>}\hn_i \hn_j ,
\end{align}
where $\hc_{i,\sigma}^{(\+)}$ annihilates (creates) an electron with spin $\sigma = \{\up,\dn\}$ on lattice site $i$, and $\<i,j\>$ represents pairs of nearest-neighbor sites. The density operator is denoted by  $\hn_{i}=\hn_{i\up}+\hn_{i\dn}$ with $\hn_{i\sigma}=\hc_{i\sigma}^{\+}\hc_{i\sigma}$ and the chemical potential is given by $\mu$.
For the half-filled case we set $\mu=U/2+4V$.
If $U$ is the dominant energy scale, the system is in a Mott insulating phase with predominantly singly occupied sites and strong antiferromagnetic correlations. A large $V$ favors doubly occupied and empty sites and leads to a different type of insulator with strong charge order tendencies. We suppress long-range order in our calculations.

To compute the dynamics of the system described by Hamiltonian~\eqref{eq:EHM} we use a nonequilibrium version of the dynamical cluster approximation (DCA)~\cite{tsuji2014,eckstein2016}, which enforces translational invariance on the cluster. In this formalism, all the sites of the cluster are hybridized with a self-consistently determined bath.
The cluster Hamiltonian is given by
\bsplit{
	H^c(t)=&-\sum_{\<i,j\>\sigma}\left[t^{c}_{i,j}(t) d_{i\sigma}^\+d_{j\sigma}+h.c.\right] +U\sum_{i}  n_{i\up}n_{i\dn} \\
	&+\sum_{{\<i,j\>, \sigma\sigma'}}  V_{i,j}^c n_{i\sigma}n_{j\sigma'}
	-\mu\sum_{i} n_{i},
\label{eq:Hc}
}
where $d_{i\sigma}^{(\+)}$ represents an annihilation (creation) operator on the cluster. The hopping matrix elements ($t_{i,j}^c=\frac{2}{\pi} t_h$) and inter-site interactions ($V_{i,j}^c=\frac{2}{\pi} V$) are renormalized due to the periodic boundary conditions.
This formalism allows us to treat the short-range correlations within the periodized cluster exactly, whereas the long-range correlations are described on the mean-field level. In this work, we use a $2\times2$ cluster with periodic boundary conditions, which gives $N_c=4$ patches in the reciprocal space around $\vK=\{(0,0),(\pi,0),(0,\pi),(\pi,\pi)\}$. We will focus on the dynamics deep in the Mott insulator and for the solution of the embeded cluster employ the non-crossing approximation~\cite{grewe1981, coleman1984, eckstein2010} (NCA). A detailed description of the DCA formalism is given in Appendix~\ref{sec:Method}.

The photo-doping of the initially Mott insulating state is generated by a time-dependent modulation of the hopping parameter
\begin{equation}
t_h(t) = 1+\Delta t_h e^{-(t-t_0)^2/\tau^2}\sin(\omega (t-t_0)),
\label{eq_t_h}
\end{equation}
with amplitude $\Delta t_h$. The frequency $\omega$ is chosen according to the gap size or expected exciton energy, and the hopping modulation has a Gaussian envelope with a maximum at time $t_0$ and a full width at half maximum $\tau$.
We choose to excite the system by a hopping modulation rather than by an electric field, because a gauge invariant formulation of DCA with electromagnetic fields is subtle (even to describe linear response, vertex corrections beyond the straightforward application of the DCA formalism must be included~\cite{lin2009}).
We do not expect that the precise mechanism by which the doublon-holon pairs are created has a qualitative effect on the discussed results.

\subsection{Observables}
In this section we define observables which are useful to trace the temporal evolution of the excitonic correlations.

\subsubsection{Double occupation}

The double occupation $D(t)$ is defined in real space as
\beq{
\label{eq:D}
	D(t) = \frac{1}{N_c}\sum_{i=1}^{N_c} \<\hn_{i\up} \hn_{i\dn}\>(t).
}
The cluster sites $i$ are numbered in a clockwise fashion (modulo 4). In practice, within the DCA formalism it is more convenient to measure $D(t)$ in momentum-space. The corresponding expression is given in Appendix~\ref{sSec:CorrFct}.

\subsubsection{Nonlocal correlation functions}

To investigate the temporal evolution of the excitonic correlations, i.e., to measure the nearest-neighbor doublon-holon pairs on the cluster, we define the following correlation function:
\begin{equation}
\label{eq:Pexc}
	\begin{split}
		\Pexc (t)&= \sum_{i} \<(\hD_i \hh_{i+1}+\hD_{i+1} \hh_{i})\times \right.\\
		&\times\left. (\hn_{i+2}-2\hD_{i+2})(\hn_{i+3}-2\hD_{i+3})\>(t),
	\end{split}
\end{equation}
where $\hD_i = \hn_{i\up}\hn_{i\dn}$ is the double occupancy operator,
$\hh_i=(1-\hn_{i\up})(1-\hn_{i\dn})$ is the hole number operator, and $\hn_i-2\hD_i$ measures singly occupied sites.
This correlation function detects the presence of a single exciton on nearest-neighbor sites under the constrain that all other sites are singly occupied. The constraint eliminates contributions from other configurations of doublon-holon pairs on the cluster, such as bi-excitons.

Analogously, to capture the dynamics of a doublon-holon pair on the diagonals of the cluster, we define
\begin{equation}
\label{eq:PexcNNN}
	\begin{split}
		\Pexc^\text{NNN} (t)&= \sum_{i} \<\hD_i (\hn_{i+1}-2\hD_{i+1})  \hh_{i+2} \times \right.\\
		&\times\left.(\hn_{i+3}-2\hD_{i+3})\>(t).
	\end{split}
\end{equation}

Finally, we introduce the bi-excitonic correlation function, which measures two doublons and two holons on the diagonals of the cluster. This function is defined as:
\begin{equation}
\label{eq:Pbexc}
	\Pbexc (t) = \frac{1}{2}\sum_{i} \<\hD_i \hh_{i+1} \hD_{i+2} \hh_{i+3}\>(t),
\end{equation}

Similar to the double occupation $D(t)$ it is useful to measure also $\Pexc$, $\Pexc^\text{NNN}$,  and $\Pbexc$ in momentum space.

\section{Results}
\label{sec:Results}

\subsection{Equilibrium}
\label{sSec:Eq}

Let us first discuss some equilibrium results of the EHM obtained by the DCA method with the NCA impurity solver. We choose the hopping parameter $t_h=1$ as the unit of energy and the on-site interaction strength $U=25$, which leads to a Mott gap whose size is about twice the width of the Hubbard band (see below). This choice is appropriate for certain Mott insulators, such as transition metal monoxides~\cite{zhang2019}. The large gap size allows us to study set-ups where the excitons lie within the continuum of doublon-holon excitations (e. g. for $V=3$), as well as set-ups with excitons in the Mott gap (e. g. for $V=6$), while avoiding charge order.  Since $V/U \lesssim 1/4$ both choices of $V$ parameters are resonable and within the range of ab-initio predictions for transition metals and perovskite compounds~\cite{miyake2008}.
The temperature is set to $T=0.1$, unless otherwise specified.

\begin{figure}[!t]
\centering
\includegraphics[width=0.95\columnwidth]{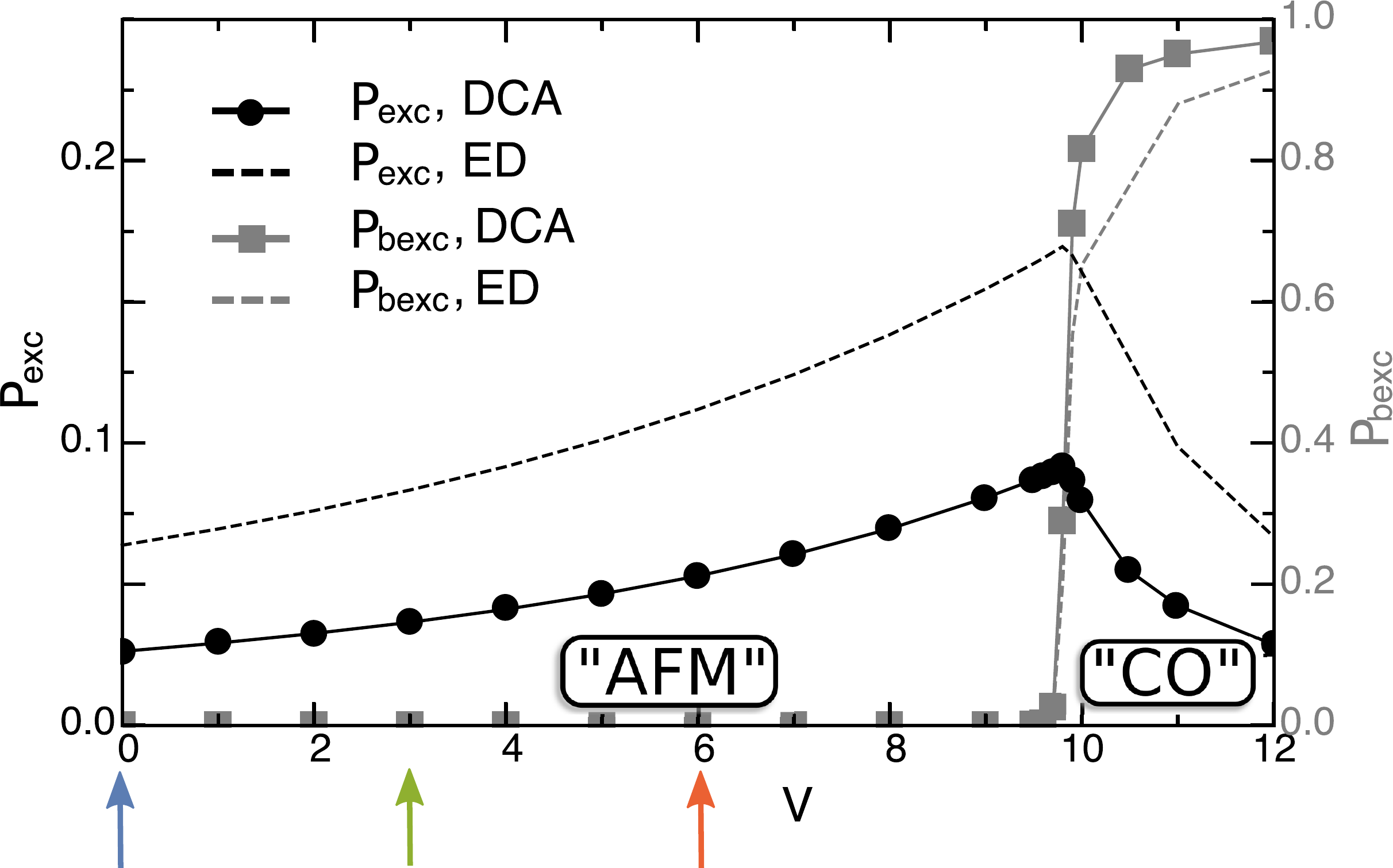}
\caption{(color online) Nearest-neighbor excitonic $\Pexc$ (black dots, Eq.~\eqref{eq:Pexc}) and bi-excitonic $\Pbexc$ (gray squares, Eq.~\eqref{eq:Pbexc}) correlation functions versus the nearest neighbor interaction $V$ at half-filling for $U=25$ and temperature $T=0.1$. The solid lines show the DCA results and the dashed lines ED data for an isolated plaquette with a renormalized $V^c$ at $T=0.1$. 
The arrows indicate the $V$ values for which results are plotted with the same colors in the other figures.
}
\label{fig:PexcV}
\end{figure}

\subsubsection{Correlation functions at half-filling}

In Fig.~\ref{fig:PexcV} we plot the excitonic correlation function for nearest-neighbor doublon-holon pairs~\eqref{eq:Pexc} as a function of the nearest-neighbor interaction strength $V$ (black solid line).
Already for $V=0$, $\Pexc$ is nonzero due to virtual hopping.
For small to intermediate nearest-neighbor interaction ($V\lesssim9.7$) $\Pexc$ shows an increase with $V$, whereas for larger $V$  the excitonic correlations are suppressed. In contrast, the bi-excitonic correlation function $\Pbexc$ [Eq.~\eqref{eq:Pbexc}] takes small values for $V\lesssim9.7$, and is strongly enhanced for larger nearest-neighbor interactions.  This behavior can be explained by a transition to a ``charge order" (CO) type insulating state (as also indicated by other correlation functions, see Appendix~\ref{sec:appEq}), even in these simulations with enforced translational symmetry. The excitonic correlation function for the doublon and holon pairs on the diagonals of the cluster remains tiny ($<10^{-3}$) for the same values of the nearest neighbor interaction $V$.

For comparison, we also performed calculations  of $\Pexc$ and $\Pbexc$
using the {\tt SNEG}~\cite{sneg2011} package, which is based on the
exact diagonalization (ED) technique. Here, we considered the extended Hubbard model Eq.~\eqref{eq:Hc} on an isolated $2\times2$ cluster with periodic boundary conditions. Due to the enforced periodization in DCA (see Appendix~\ref{sec:Method}), a proper comparison with ED requires an on-site interaction $U$ and a renormalized nearest-neighbor interaction $V^c=\frac{2}{\pi}V.$ (Since in our DCA calculations
there is no bosonic self-consistency leading to a screened $U$ the relevant ratio of interaction parameters is $U/V^c$.)
As one can see in Fig.~\ref{fig:PexcV} there is a qualitative agreement between the DCA and ED results for the correlation function $\Pexc$, and an almost quantitative agreement for the transition point to the ``CO" regime. The quantitative difference in $\Pexc$ originates from the coupling to the fermionic bath, which is included in the DCA calculations. The hybridization with the bath leads to an increase in the population of non-half-filled states and as a result to smaller values of $\Pexc$ compared to the ED results.

The energy cost associated with the formation of an exciton on a $2\times2$ plaquette in atomic limit is $U-2V^c$, with a correction of order ${\cal O} (t_h^2/(U-2V^c))$ due to the small $t_h$. In the simpler case of an isolated periodized dimer this energy can be obtained analytically (for details see Appendix~\ref{Ssec:appEexc}):
\begin{equation}
\label{eq:Eexc}
	E_\text{exc}\approx U-2V^c+\frac{16t_h^2}{(U-2V^c)}.
\end{equation}
In following we will use Eq.~\eqref{eq:Eexc} as a rough estimate of the exciton binding energy on a 2D plaquette.
On the other hand, the splitting between the Hubbard bands is approximately $\Delta_\text{Mott}= U-W$, with $W$ the bandwidth. The exciton is expected to lie within the doublon-holon continuum for $E_\text{exc}>\Delta_\text{Mott}$.
For $U=25$ this means that the exciton lies inside the continuum for $V\lesssim 4.9$ ($V^c\lesssim 3.1$) . In order to understand the role of excitons in the nonequilibrium dynamics of Mott insulators we therefore consider in the following two representative cases: (i) $V=3$, where the exciton energy is within the doublon-holon continuum, and (ii) $V=6$, where the exciton appears within the Mott gap.

\subsubsection{Local spectral function}

\begin{figure}[!t]
\centering
	\includegraphics[width=0.95\columnwidth]
{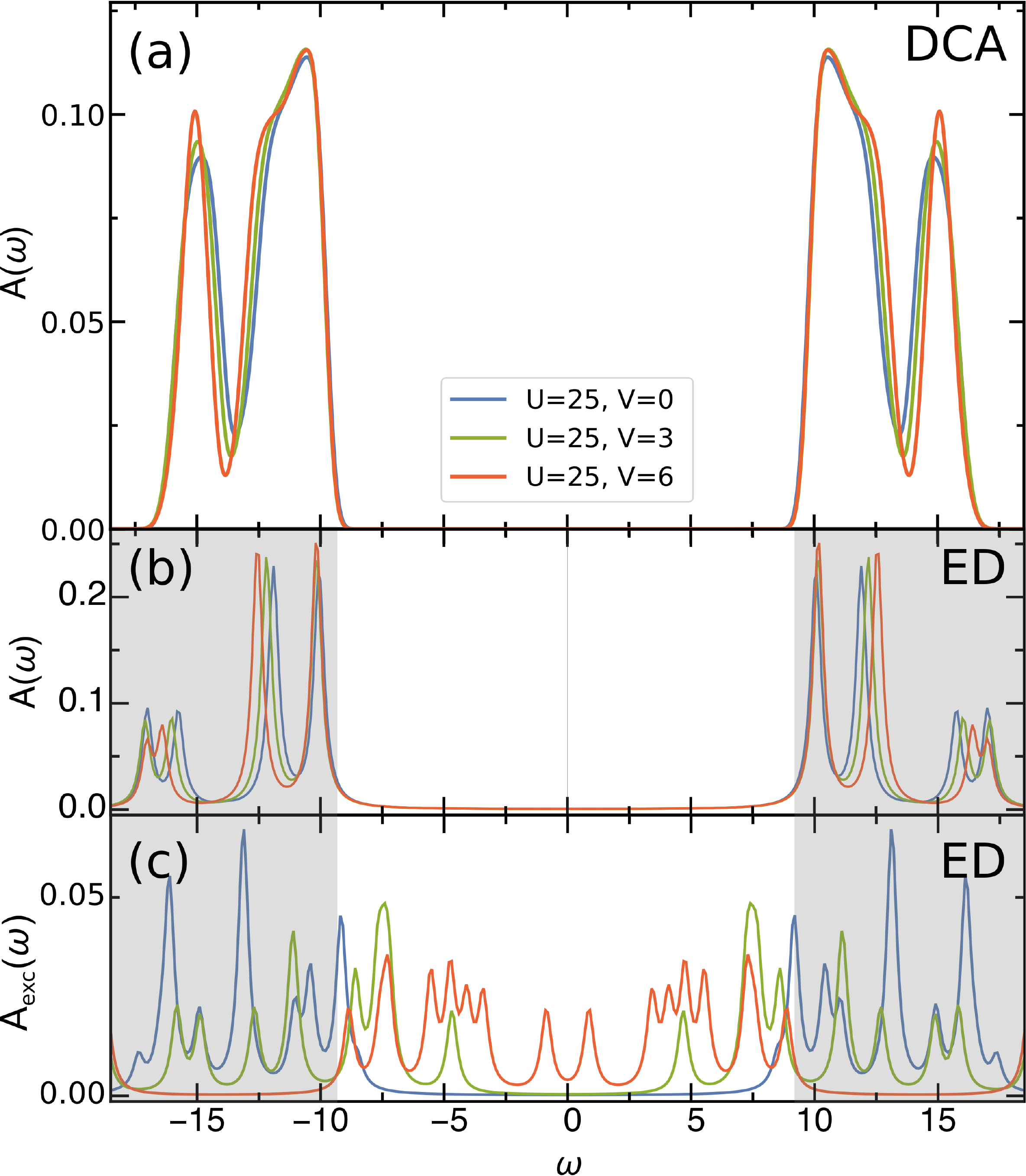}
\caption{(color online) Equilibrium spectral function $A(\omega)$ for different nearest-neighbor interactions $V$ (color-coded) obtained using (a) DCA and (b) ED. The calculations are done for $U=25$ at temperature $T=0.1$.
(c) The sum of the dominant contributions from $\vK=(0,0)$ and $\vK=(\pi,\pi)$ to the local
spectral function. The calculations are done with ED for an isolated cluster with $U=25$ and $V^c=\tfrac{2}{\pi} V$ at temperature $\beta=0$ (2x2 cluster) and the spectra are multiplied by a factor $10^{2}$. Different colors correspond to different values of the nearest-neighbor interaction $V$ (see labels in panel (a)).
}
\label{fig:Aw}
\end{figure}

We calculate the spectral function $A(\omega)$ from the retarded component of the local Green function,
\beq{
	A(\omega)=-\frac{1}{\pi} \mathrm{Im} G^R(\omega),
}
performing the Fourier transformation of $G=1/N_c\sum_\vK G_\vK$ within a time window of length $t_\mathrm{max}=10$.
The results for different nearest-neighbor interaction strengths $V$ are shown in Fig.~\ref{fig:Aw}a.
The model with $V=0$ yields two Hubbard bands with a width $W \approx 8$, which are separated by a gap  $\Delta_\text{Mott}\approx 19.6$. Increasing the nearest neighbor interaction strength $V$ within the ``AFM" region leads
to a redistribution of the spectral weight within the bands.
For $V\gtrsim 9.7$, after the transition to the ``CO" insulating state, the gap between the bands increases with $V$ (not shown). A similar behavior is found in the ED calculations for an isolated cluster
with renormalized $V^c$ (see Fig.~\ref{fig:Aw}(b)), where we find a gap of
similar size in the ``AFM" parameter region and an increase of the gap size in the ``CO" region (not shown).

In order to analyse the spectral signatures of excitonic states, we calculate using ED their contributions to $A(\omega)$ and show the results in Fig.~\ref{fig:Aw}(c). 

From the Lehmann representation it follows that the contribution of an eigenstate $\ket{\psi}$ with eigenenergy $\epsilon$ to the $\vK$-resolved
spectral function is
\begin{equation}
\label{eq:IwLehmann}
	\begin{split}
	A_{\ket{\psi}}(\omega,\vK) =& -\mathrm{Im}\Bigg\{\frac{1}{Z}\sum_{n}\frac{\left|\bra{n} c^\+_\vK\ket{\psi}\right|^2}{\omega - (\epsilon_n-\epsilon)+i\eta}\times\\
	&\times\left(e^{-\beta\epsilon_n}+e^{-\beta\epsilon}\right)\Bigg\},
	\end{split}
\end{equation}
where the $\ket{n}$ denote the eigenstates of the Hamiltonian with the corresponding eigenvalues $\epsilon_n$, and $\beta=1/T$ is the inverse temperature.
For $\ket{\psi}$ we consider the four eigenstates of the system with the strongest excitonic correlations ($\bra{\psi}\Pexc\ket{\psi}\approx1$).
These excitonic-like states contribute negligibly little to the total spectral function in a low-temperature equilibrium state. Since we are only interested in the positions of the corresponding peaks we set $\beta=0$ in the ED spectra of Fig.~\ref{fig:Aw}(c).
For the broadening of the peaks we use $\eta=0.25$.

For each value of the nearest-neighbor interaction the cluster-momentum $\vK$-resolved spectrum $A_\text{exc}(\omega,\vK)$ exhibits several distinct peaks and in Fig.~\ref{fig:Aw}(c) we present the sum of the dominant contributions
\begin{equation}
\label{eq:AwK}
	A_\text{exc}(\omega)=\sum_{\ket{\psi} \text{with}\atop \bra{\psi}\Pexc\ket{\psi}\approx 1}\left\{A_{\ket{\psi}}(\omega,(0,0))+A_{\ket{\psi}}(\omega,(\pi,\pi))\right\}.
\end{equation}
 In the ``AFM" regime, the dominant excitonic peaks 
shift towards frequency $\omega=0$ with increasing $V$.
A comparison between the spectral functions obtained from DCA and from ED shows that the high-energy exciton peaks appear in the energy region of the Hubbard bands. However, for large enough $V$, the dominant low-energy peaks in the excitonic contribution to the spectrum lie within the Mott-gap,
and, as we show later, can become visible in the photo-doped state.

\subsection{Non-Equilibrium}
\label{sSec:NonEq}

In this section, we study the properties of a photo-doped Mott insulating state described by the EHM ($U=25$, $V=0,3,6$). The photo-excitation is performed
by a hopping modulation
with a Gaussian envelope, Eq.~(\ref{eq_t_h}), with frequency $\omega=20$ centered at $t_0=0.45$, full width at half maximum $\tau=0.3$, and amplitude $\Delta t_h=0.75$. This pulse excites electrons across the Mott gap and creates long-lived doublons and holons.
Energy is measured in units of $t_h$ and time in units of $\hbar/t_h$.

\begin{figure}[!b]
\centering
	\includegraphics[width=1.0\columnwidth]
{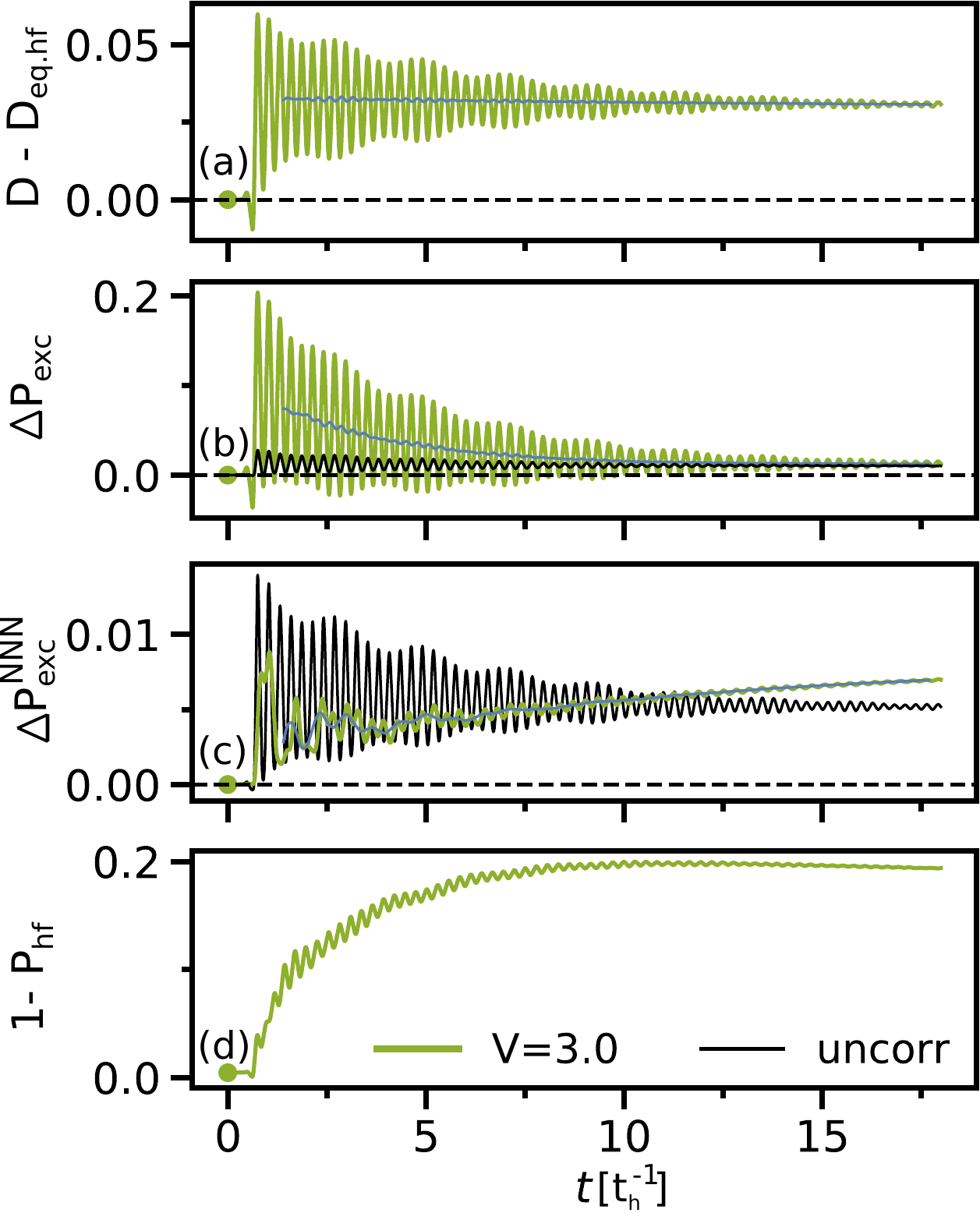}
\caption{(color online) Simulation results for $U=25$, $V=3$ and initial temperature $T=0.1$ showing the temporal evolution of the change in the correlation functions after a hopping modulation  with $\Delta t_h=0.75$, and $\omega=20.0$. (a) Double occupancy measured relative to the equilibrium value ($D_\mathrm{eq.,hf}$), (b) photo-induced doublons and holons on nearest neighbor sites ($\Delta\Pexc$),~(c) doublon-holon pairs on a diagonal of the cluster ($\Delta\Pexc^\mathrm{NNN}$). The corresponding probability of non-half-filled plaquette states is shown in panel (d).
The blue solid line in (a) and (b) shows the running average over one oscillation. 
The black solid lines represent the probabilities for uncorrelated doublons and holons on neigboring sites (Eq.~\eqref{eq:uncorrPexc}) and on the diagonals of the cluster (Eq.~\eqref{eq:uncorrPexcNNN}).
}
\label{fig:tCorrV03}
\end{figure}

\subsubsection{Short-lived excitons}
\label{sSec:ShortLivedExc}

In this subsection, we focus on a system for which the excitons lie inside the continuum of doublon-holon excitations ($V\lesssim 4.9$). The photo-excitation leads to a strong increase in the double occupation (see Fig.~\ref{fig:tCorrV03}(a)) with subsequent strong but damped oscillations.

\begin{figure}[!b]
\centering
	\includegraphics[width=0.95\columnwidth]
{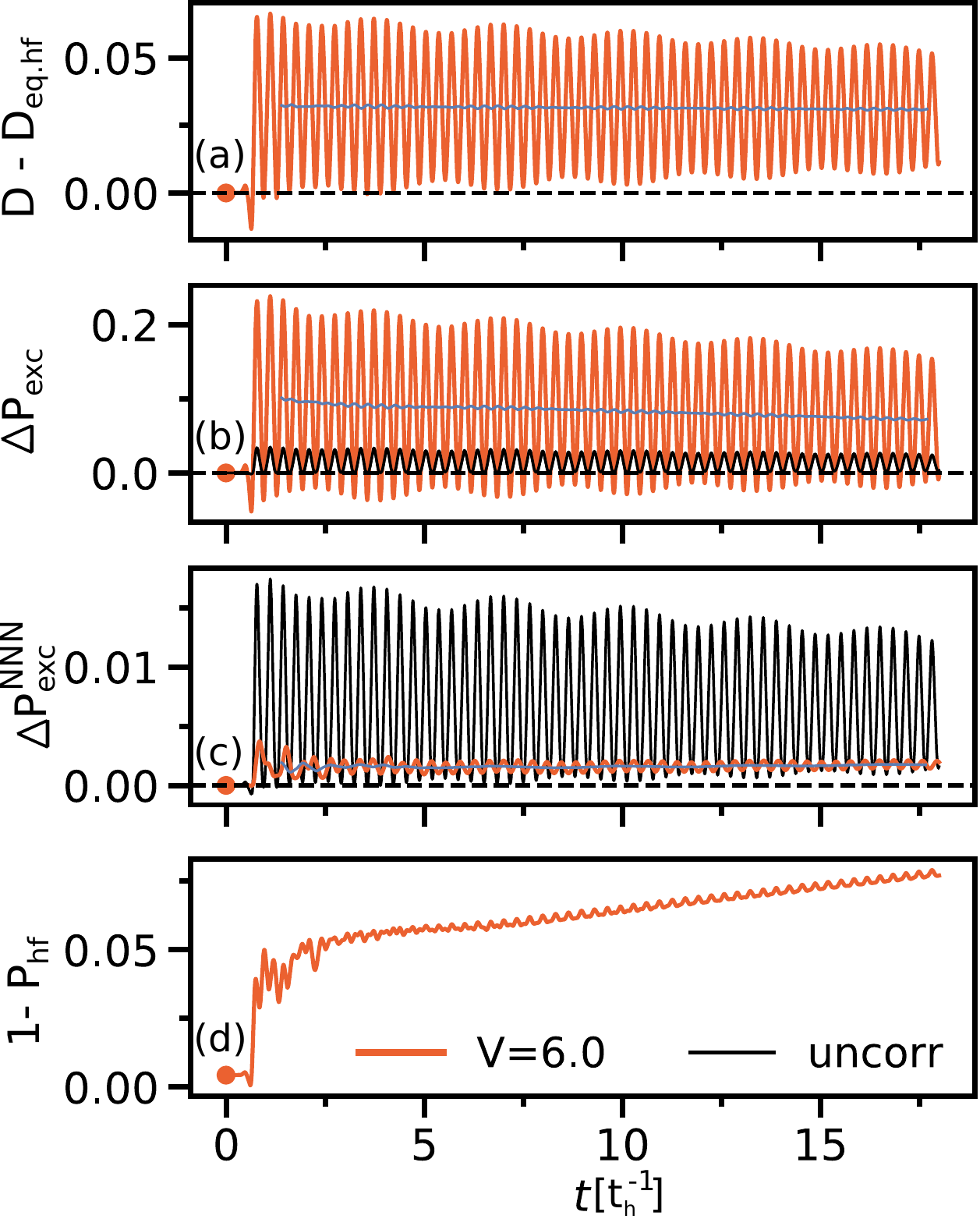}
\caption{(color online) Simulation results for $U=25$, $V=6.0$ and initial temperature $T=0.1$ showing the temporal evolution of the change in the correlation functions after a hopping modulation with $\Delta t_h=0.75$, and $\omega=20.0$. (a) Double occupancy measured relative to the equilibrium value ($D_\mathrm{eq.,hf}$), (b) photo-induced doublons and holons on nearest neighbor sites ($\Delta\Pexc$), (c) doublon-holon pairs on a diagonal of the cluster ($\Pexc^\text{NNN}$), and (d) the probability of non-half-filled plaquette states $1-\Phf$. The black solid lines represent the changes in the probabilities for uncorrelated doublons and holons on neighboring sites (Eq.~\eqref{eq:uncorrPexc}) and on the diagonals of the cluster (Eq.~\eqref{eq:uncorrPexcNNN}). The blue solid line represents the running average over one oscillation.
}
\label{fig:tCorrV06}
\end{figure}

Let us start with the temporal evolution of the doublon-holon correlation functions after a hopping modulation for $V=3$. In order to exclude the virtual contributions to $\Pexc(t)$ and $\Pexc^\text{NNN}(t)$ we plot in Fig~\ref{fig:tCorrV03}(b) and (c) the changes of these functions with respect to their equilibrium values.
As one can see from Fig.~\ref{fig:tCorrV03}(b),
the high-frequency excitation  (which lasts up to $t\approx 1.2$) results in a significant initial increase of the correlations $\Pexc$, and a subsequent decay within a relaxation time of $\approx 3$ (extracted from a fit to the running average in panel (b)).
Superimposed on this evolution are strong oscillations, whose frequency
 is essentially independent of the excitation strength
and is $\approx U-2V^c$. Therefore, these oscillations can be interpreted as coherence between the photo-induced exciton and the ground state. Later, we take this oscillation frequency as a measure for the exciton binding energy.
While the
value of P$_\mathrm{exc}$ at longer times is enhanced compared to the initial equilibrium value this does not automatically imply the existence of long-lived photo-induced doublon-holon pairs.
As a reference value, we calculate the probability of  uncorrelated doublons and holons on neighboring sites:
\begin{equation}
	\Pexc^\text{uncorr}(t) = 8[D(t)(1-2 D(t))]^2\ ,
	\label{eq:uncorrPexc}
\end{equation}
where we have used that in the half-filled system the expectation value for doublons is the same as that for holons.
The result is shown in Fig~\ref{fig:tCorrV03}(a) by the black solid line (note that for both $\Pexc(t)$ and $\Pexc^\text{uncorr}(t)$ we plot the difference to the initial value). The comparison of $\Delta\Pexc(t)$ with $\Delta\Pexc^\text{uncorr}$ indicates that the pulse initially generates bound nearest-neigbor doublon-holon pairs, which subsequently decay almost completely into uncorrelated doublons and holons.

The correlation function which measures 
doublon-holon pairs on the diagonals of the cluster is initially tiny (not shown).
There is a systematic enhancement after the excitation over the Mott gap with weak oscillations, but the value remains small, see Fig.~\ref{fig:tCorrV03}(b). The probability for uncorrelated doublons and holons on the diagonals of the cluster can be estimated in analogy to Eq.~\eqref{eq:uncorrPexc}:
\begin{equation}
	\Pexc^\text{uncorr, NNN}(t) = 4[D(t)(1-2 D(t))]^2\ .
	\label{eq:uncorrPexcNNN}
\end{equation}
The comparison of $\Delta\Pexc^\text{uncorr, NNN} (t)$ with $\Delta\Pexc^\text{NNN} (t)$ (see Fig.~\ref{fig:tCorrV03}(c)) shows that 
the small increase in $\Pexc^\text{NNN} (t)$ with respect to its equilibrium value is to a large extent explained by the increase in $\Pexc^\text{uncorr, NNN} (t)$. The bi-excitonic correlation function $\Pbexc$ 
on the other hand shows an increase which is not primarily due to an increase in uncorrelated doublon-holon pairs,
see Fig.~\ref{fig:tPbexc} in Appendix~\ref{sec:apptCorr}. However, $\Pbexc$ takes very small values, so that the photo-generation of biexcitons is essentially negligible in our model. A much more significant effect is the increase in the probability of the non-half-filled plaquette states (see Fig.~\ref{fig:tCorrV03}(d)).
Hence, we interpret the data shown in Fig.~\ref{fig:tCorrV03} as follows: A strong photo-doping across the gap creates a large density of doublons and holons, initially on nearest-neighbor sites. Within a time of $\approx 3$ these meta-stable excitons separate, and the corresponding unbound doublons and holons result in an increase in the probability of plaquette configurations with 5 or 3 electrons.

\begin{figure}[!b]
\centering
	\includegraphics[width=0.75\columnwidth]
{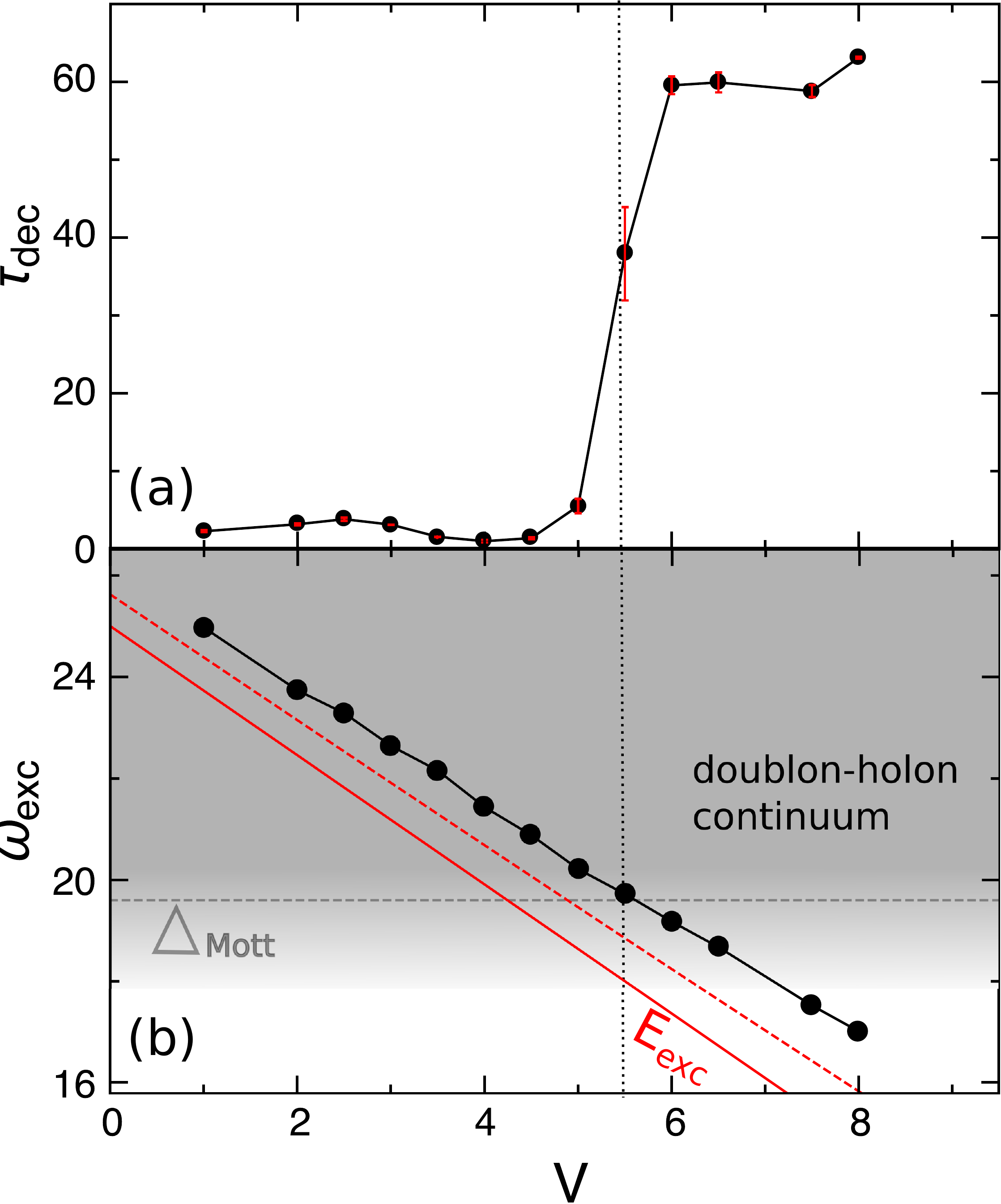}
\caption{(color online) (a) Relaxation time $\tau_{\text{dec}}$ of $\Pexc$ versus nearest neighbor interaction $V$. (b) Measured exciton energy $\omega_\text{exc}$ as a function of $V$. The grey dashed line shows the Mott gap $\Delta_{\text{Mott}}$ for $U=25$ in the ``AFM" regime. The red lines illustrate the energy cost for forming an exciton on a plaquette in the atomic limit (solid) and on an isolated dimer (dashed), as estimated by Eq.~\eqref{eq:Eexc}. The vertical dotted line 
indicates the value of $V$ where the exciton shifts out of the doublon-holon continuum.
}
\label{fig:TauV}
\end{figure}

\subsubsection{Long-lived excitons}

In this subsection, we consider the second set-up with excitons lying in the Mott gap region ($V\gtrsim 4.9$).

Figure~\ref{fig:tCorrV06} plots the temporal evolution of the double occupancy and the correlation functions after the photo-excitation in the model with $V=6$. The photo-excitation produces similar enhancement in $D-D_\mathrm{eq.,hf}$ (see Fig.~\ref{fig:tCorrV06}(a)) as in the case with $V=3$ (Fig.~\ref{fig:tCorrV03}(a)). However, the oscillations in the double occupancy are only weakly damped.
Now, we focus on the changes in the excitonic correlation function $\Delta\Pexc$, which measures the photo-induced nearest-neighbor doublon-holon pairs. The result is shown in Fig.~\ref{fig:tCorrV06}(b) by the solid red line. In contrast  to the case of short-lived excitonic states within the doublon-holon continuum (Fig.~\ref{fig:tCorrV03} (b)), we find here a strong enhancement of $\Pexc$ with respect to its equilibrium value and long-lived pronounced oscillations. The probability for uncorrelated doublons and holons (black solid line) remains very low in this case, which indicates that long-lived nearest-neighbor excitons are photo-induced.
The evolution of the correlation function for doublon-holon pairs on the diagonals of the cluster is illustrated in Fig.~\ref{fig:tCorrV06}(c). Here, we observe a tiny enhancement of $\Pexc^\text{NNN}$ with respect to its equilibrium value with subsequent oscillations. The value of $\Delta\Pexc^\text{NNN}$ is smaller than $\Delta\Pexc^\text{uncorr, NNN}$ (black solid line), indicating a suppression of the next-nearest neighbor correlations by photo-excitation.
While long-lived bi-excitons are photo-generated (Fig.~\ref{fig:tPbexc}), their density remains low.
Finally, we only find a small increase in the probability of non-half-filled plaquette states (see Fig.~\ref{fig:tCorrV06}(d)), which is also consistent with  the creation of long-lived nearest-neighbor excitons by the photo-excitation.

\begin{figure}[!t]
\centering
	\includegraphics[width=1.0\columnwidth]{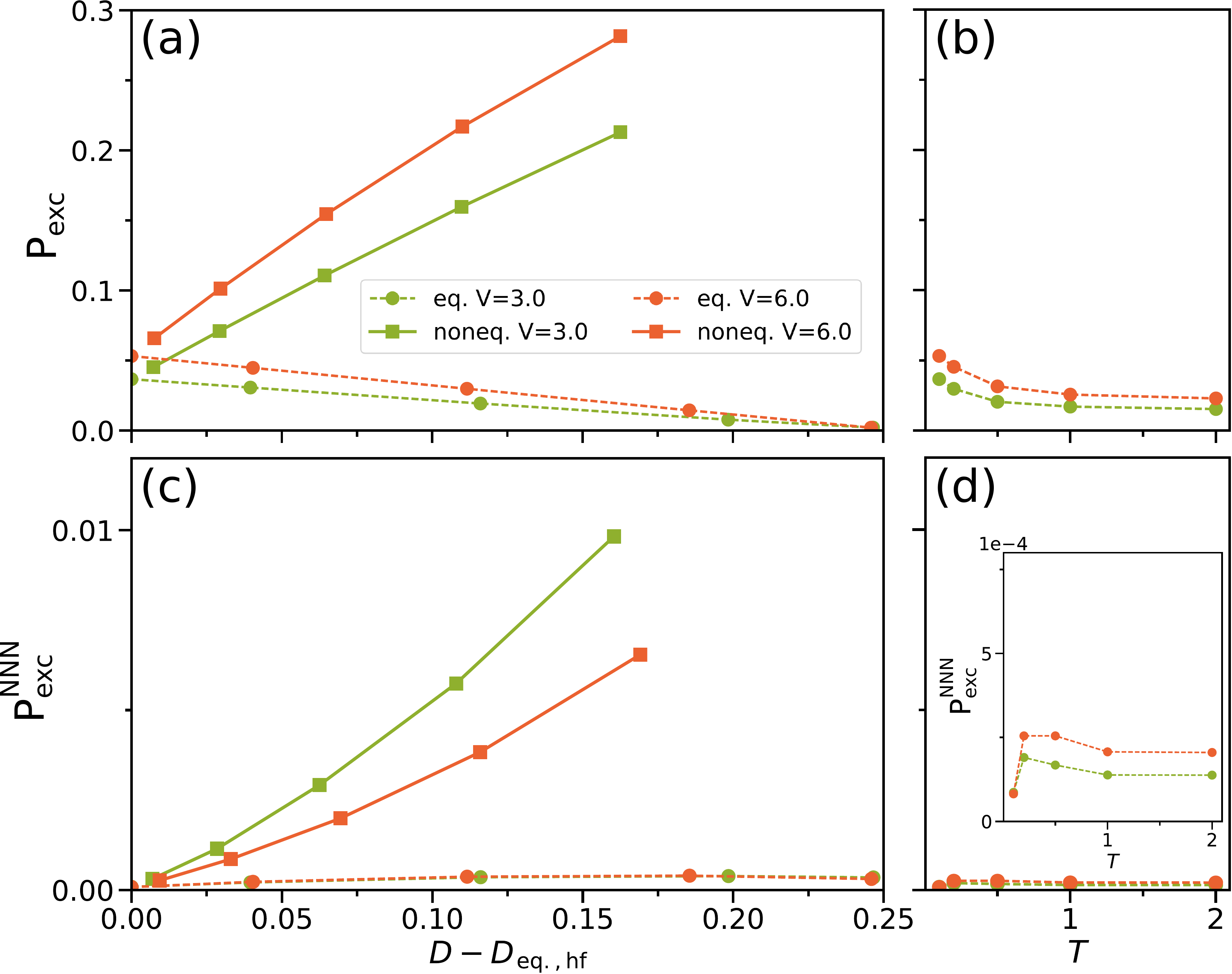}
\caption{(color online)  Comparison of photo-doped and chemically doped systems: (a) $\Pexc$ and (c) $\Pexc^\text{NNN}$
vs. double occupancy measured relative to the equilibrium half-field ($D_\mathrm{eq., hf}$) case. The calculations are done for $U=25$ at temperature $T=0.1$. Different colors correspond to different values of the nearest-neighbor interaction $V$.
Squares indicate the results of the nonequilibrium calculations, measured directly after the pulse, whereas dots represent equilibrium results for chemically doped systems. Panels (b) and (d)
show the temperature dependence of these correlation functions. These results were obtained in equilibrium at half-filling using the DCA method and confirmed qualitatively by ED calculations (not shown).
}
\label{fig:Pnoneq}
\end{figure}

In order to extract the life-time of the excitons for different values of the nearest neighbor interaction $V$, we fit the running average of the excitonic correlation function $\bar{P}_\text{exc}(t)$ in the time interval $t\in \left[2.2, 16\right]$ with a single exponential function:
\begin{equation}
	\bar{P}_\text{exc}(t) = A +  B\cdot\exp(-t/\tau_\text{dec}),
\end{equation}
where $A$ and $B$ are fitting parameters, and $\tau_\text{dec}$ denotes the relaxation time. In Fig.~\ref{fig:TauV}(a) we plot the extracted $\tau_\text{dec}$ as a function of the nearest-neighbor interaction $V$. As one can see, the relaxation time for $V\lesssim 5$ is short, which indicates a fast dissociation of the photo-excited excitons. Around $V\approx 5.5$ one observes a strong increase of $\tau_\text{dec}$, which signals the creation of long-lived excitonic states for larger $V$.
(While a $\tau_\text{dec}\approx 60$ extracted from a fit over $t\lesssim 15$ has a large uncertainty, which is not properly reflected in the error bars, the dramatic increase in $\tau_\text{dec}$ is an unambiguous result. 
This increase is the result of a shift of the excitonic states from the doublon-holon continuum into the Mott-gap, where the excitonic states are  isolated. To illustrate this point we plot with black dots in Fig.~\ref{fig:TauV}(b) the energy $\omega_\text{exc}$ of the excitonic states, measured as the frequency of the oscillations in $\Pexc(t)$, versus $V$. While for $V\lesssim 5.5$ the excitons are lying within the doublon-holon continuum, for $V\gtrsim 5.5$ these states appear within the Mott gap.
We also plot by a red dashed line the simple dimer based estimate for $E_\text{exc}$ (Eq.~(\ref{eq:Eexc})), which gives a remarkably good prediction of the measured exciton energy.

\begin{figure}[!b]
\centering
	\includegraphics[width=1.0\columnwidth]
{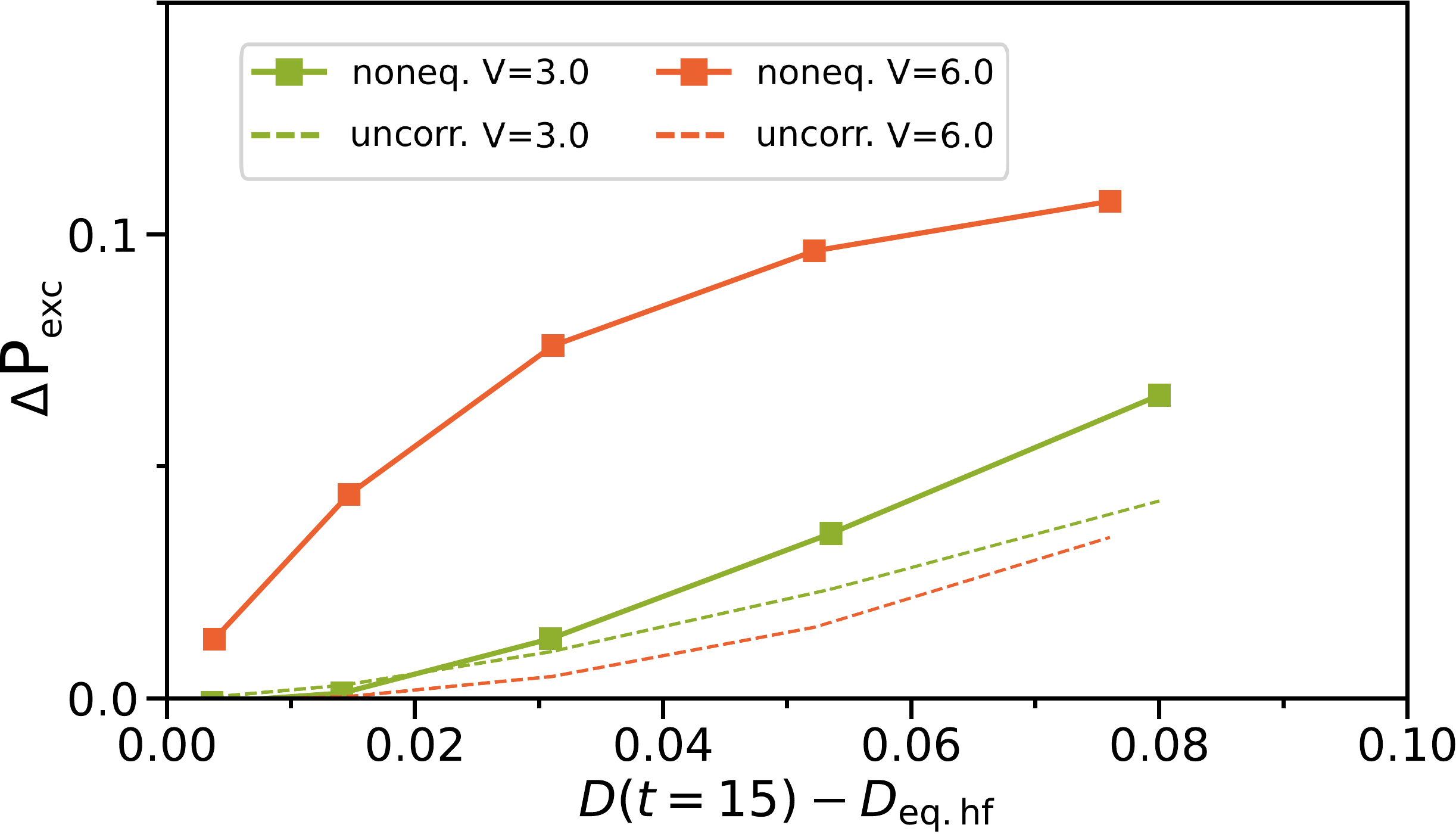}
\caption{(color online) Dependence of the photo-induced excitonic correlation function $\Delta \Pexc$ measured at $t=15$ on the
double occupancy at $t=15$ measured relative to the equilibrium half-filled ($D_\mathrm{eq.,hf}$) case.
The calculations are done at $T=0.1$ for $U=25$, $V=3$ (green squares) and $V=6$ (red squares). The dashed lines represent the change in the probability of uncorrelated doublons and holons on neigboring sites, $\Delta \Pexc^\text{uncorr}$. 
}
\label{fig:Pexcth}
\end{figure}

\begin{figure*}[!htb]
\centering
	\includegraphics[width=0.95\textwidth]
{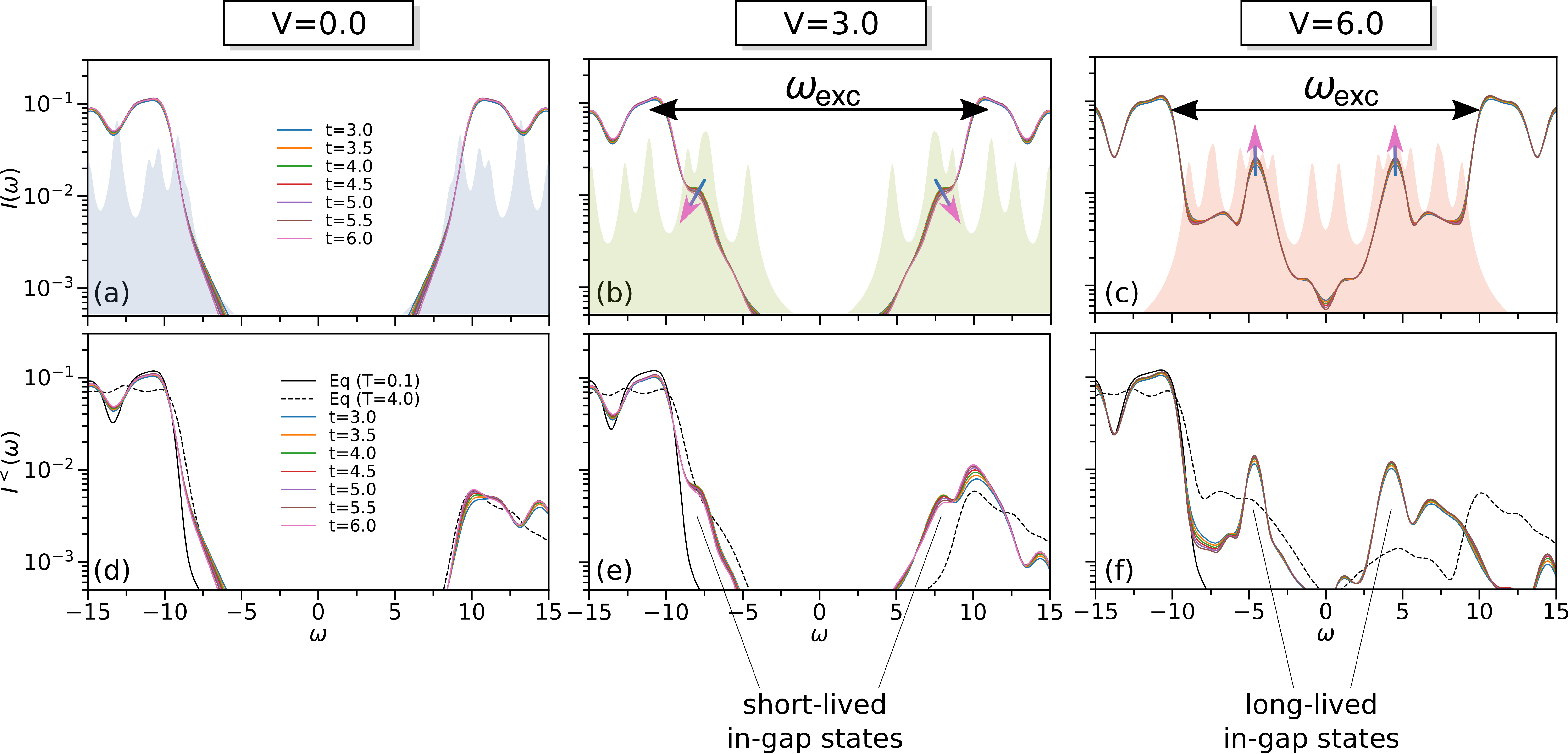}
\caption{(color online) Non-equilibrium photoexcitation spectrum calculated at different times (coloured lines) and plotted on a logarithmic scale for $U=25$, (a),(d) $V=0.0$, (b),(e) $V=3.0$, and (c),(f) $V=6.0$.
The spectrum $I(\omega)$ calculated from the retarded component of the Green's function (see Eq.~\eqref{eq:spectrum}) is shown in (a)-(c) together with the sum of the equilibrium spectral functions at $\vK=(0,0)$ and $(\pi,\pi)$ obtained from ED calculations (shadowed regions), whereas the occupation functions $I^<(\omega)$ are presented in panels (d)-(f). (The shadowed ED spectra in (a)-(c) are multiplied by a factor of $10^2$, similar to Fig.~\ref{fig:Aw}(c).) The black lines represent equilibrium results at $T=0.1$ (solid) and $T=4.0$ (dashed). 
}
\label{fig:IpesNonEq}
\end{figure*}

\subsubsection{Photo-doping dependence}

In Fig.~\ref{fig:Pnoneq} we analyze the dependence of the excitonic correlation functions on chemical doping (dots) and photo-doping (squares),
respectively. For this we plot $\Pexc$ as a function of the change in the double occupancy relative to the equilibrium half-filled value $D_\mathrm{eq., hf}$. In order to take into account the creation of both doublons and holons by the photo-doping, we multiplied $D-D_\mathrm{eq.,hf}$ by a factor of 2 in the photo-doped case. For $D$ we take the running average value (within $\Delta t\approx2\pi/\omega_\mathrm{exc}$) measured directly after the pulse. We focus first on the excitonic correlation function $\Pexc$ for neighboring doublon-holon pairs, which is also estimated from the running average measured
directly after the pulse (solid lines with squares in Fig.~\ref{fig:Pnoneq}(a)).
Clearly, the photo-doping results in enhanced initial excitonic correlations while chemical doping suppresses $\Pexc$ (dashed lines with circles).
This cannot be explained by a simple heating effect, since the latter results in the opposite trend, see Fig.~\ref{fig:Pnoneq}(b). 

Now we turn to the excitonic correlation function $\Pexc^\text{NNN}$ for a doublon-holon pair on a diagonal of the cluster, shown in Fig.~\ref{fig:Pnoneq}(c), where we plot the running average value (within $\Delta t\approx0.36$) of $\Pexc^\text{NNN}$ measured directly after the pulse in the nonequilibrium case (squares). Similar to the case of the excitonic correlation function for neighboring doublon-holon pairs we find an enhancement of the excitonic correlations along the diagonals of the cluster after the photo-excitation for different values of $V$, even though the values are an order of magnitude smaller. We note, however, that the next-nearest neighbor correlation gets suppressed by increasing $V$ in contrast to the nearest neighbor correlations.
While the temperature dependence of $\Pexc^\text{NNN}$ shows a tiny increase even for $V=3$ (see Fig.~\ref{fig:Pnoneq}(d)), the effect is too small to explain the much more significant increase observed in the photo-doped states.
Hence, also the relatively strong enhancement of the excitonic correlations along the diagonals of the cluster observed for different values of $V$ in Fig.~\ref{fig:Pnoneq}(c) is a property of the photo-doped non-thermal state.

Finally, we analyze the long-time behavior of the excitonic correlations. For this we plot in Fig.~\ref{fig:Pexcth} the changes in the running average value at $t=15$ as a function of the change in the double occupancy relative to the equilibrium half-filled value. For $D$ we also take the running average value measured at $t=15$.
As one can see, $\Delta\Pexc$ shows an enhancement with the photo-doping (solid lines with squares).
However, for very strong excitations the value of $\Delta\Pexc$ starts to show a saturation effect, while the density of uncorrelated doublon-holon pairs grows  (see dashed lines in Fig.~\ref{fig:Pexcth}).
This shows that in the strong excitation regime, a larger fraction of the excitons separates into unbound doublons and holons.

\subsubsection{Non-equilibrium spectral functions}

To gain additional insights into the nonequilibrium properties of the photo-excited extended Hubbard model
we calculate the time-dependent spectral functions~\cite{freericks09}. First, we focus on the spectrum calculated from the retarded component $G^R$ of the local Green's function:
\begin{align}
\label{eq:spectrum}
	I(\omega, t_p)=&-\mathrm{Im}\int d\bar{t} d\bar{t'} e^{-i\omega(\bar{t}-\bar{t'})} G^R (t_p+\bar{t}, t_p+\bar{t'}) \nonumber \\
	&\times S(\bar{t}) S(\bar{t'}), 
\end{align}
where $S(t)\propto \exp[-(t^2/(2\Delta t_\mathrm{probe}^2)]$
is the envelope of the probe pulse of length $\Delta t_\mathrm{probe}=3.0$. 
$S(t)$ is normalized in such a way that the integral~\eqref{eq:spectrum} without $G^R$ gives $1/\pi$.

In Fig.~\ref{fig:IpesNonEq}(a)-(c) we plot the results of the DCA calculations for $I(\omega, t_p)$ on a logarithmic scale for $V=0$, $V=3$, and $V=6$, respectively. The shaded part corresponds to equilibrium spectral functions from ED for the singlet excitonic states with $\vK=(0,0)$ and $\vK=(\pi, \pi)$ (c.f. Fig.~\ref{fig:Aw}(c)),
calculated according to Eqs.~\eqref{eq:IwLehmann} and \eqref{eq:AwK}.
After the photo-excitation one observes a partial filling-in of the gap for all values of the non-local interaction $V$.  In addition, for $V>0$ in-gap peaks appear in the spectra, which have energies similar to those of the excitonic peaks identified in Fig.~\ref{fig:Aw}(c).
Further, from Fig.~\ref{fig:IpesNonEq}(b) and (c) one can see that the position of the in-gap peaks depends strongly on the nearest-neighbor interaction $V$
and is consistent with the position of the dominant excitonic features identified in the ED analysis (Fig.~\ref{fig:Aw}(c)).
By increasing $V$ the peaks appear at lower frequencies.
Clear signatures of these in-gap peaks can only be observed for $V>2$. The peaks at $\omega\approx 1$ and $5$ in Fig.~\ref{fig:IpesNonEq}(c) can be associated with the removal of an electron from the exciton, see process (i) in Fig.~\ref{fig:Ipes3D}. More specifically, the peak at $\omega\approx 1$ corresponds to a final state with an odd parity and the one at $\omega \approx 5$ to a final state with an even parity. 

Here we should comment that in optical experiments light couples to an odd-parity state and even-parity states are optically forbidden on the linear response level. However, a recent ED study of the one-dimensional extended Hubbard model has revealed that after a photo-excitation one can induce transitions between these two states leading to a new low-energy peak corresponding to the energy difference between an odd-parity and even-parity state~\cite{lu2015}. Hence, the main difference between the optical response and the photo-emission signal discussed in this work is that in the latter, even and odd parity states are directly visible, while in the former, transitions between them lead to a specific type of in-gap feature.

To analyze the occupation of the states, we calculate the time-dependent photo-excitation spectrum $I^<(\omega,t)$ from the lesser component of the local Green's function $G^<(t,t')$ in a way analogous to Eq.~(\ref{eq:spectrum})~\cite{freericks09}. 
In Figs.~\ref{fig:IpesNonEq}(d)-(f) we show the simulation results on a logarithmic scale for $V=0$, $3$, $6$, respectively.
The equilibrium results at $T=0.1$ and $T=4.0$ are shown by the black solid and dashed lines, respectively.
While at low temperature only the lower Hubbard band is populated, increasing temperature leads to a partial population of the upper Hubbard band and a partial filling of the gap.
As expected, the high-frequency pulse excitation creates a nonthermal population in the upper Hubbard band at all values of $V$. In addition, for $V>0$ in-gap states with a clearly nonthermal character are populated. While for $V=3$ short-lived in-gap states (decaying within $t\approx 3$) are observed, for $V=6$ long-lived in-gap states are populated. These results are consistent with the dynamics of the correlation functions (c.f. Fig.~\ref{fig:tCorrV03}(a) and Fig.~\ref{fig:tCorrV06} (a)).
The different life time of the excitonic features is more clearly evident in Fig.~\ref{fig:Ipes3D} which plots the results for $I(\omega)$ with the $\omega$ range restricted to the gap region.

We note that for $V=6$ an excitonic state with doublon-holon pair along the diagonal of the cluster could also contribute to the peak at $\omega\approx7$, since this energy corresponds to the removal of an electron from this state.

\section{Summary and Conclusions}
\label{sec:Summary}

In this paper we investigated the effect of nonlocal interactions on the nonequilibrium states in Mott insulators after charge excitations across the Mott gap. Specifically, we considered two set-ups with  (i) excitons lying within the doublon-holon continuum ($U=25$, $V=3$) and (ii)  excitons appearing within the Mott gap ($U=25$, $V=6$).
To simulate the dynamics of strongly correlated electrons after photo-excitation in the Mott insulating regime, we applied a nonequilibrium generalization of the plaquette dynamical cluster approximation to the extended Hubbard model and used the non-crossing approximation as an impurity solver. This formalism allows to measure doublon-holon correlations on the periodized four-site cluster in nonequilibrium and to calculate the time-resolved photoemission spectrum.
Additional comparison with equilibrium calculations for an isolated four-site cluster allowed us to identify different excitonic states.

Photo-excitation above the Mott gap leads to the creation of doublons and holons that move in a half-filled Mott background.
If the exciton energy lies in the continuum of doublon-holon excitations, one observes an initial enhancement of the nearest neighbor excitonic correlations in the ``AFM" state compared to chemical doping. This cannot be attributed to a heating effect, since the latter results in the opposite behavior. However, within a time of a few inverse hoppings,  a large fraction of these nearest-neighbor doublon-holon pairs decays into uncorrelated doublons and holons.
The transient enhancement of the excitonic correlations after the photo-excitation manifests itself also in the photoemission spectrum. In particular, for modest $V$, peaks associated with the transient presence of photo-excited excitons appear near the gap edge. 
In contrast, in a model with larger $V$, where the excitons lie inside the Mott gap, photo-doping results in a significant enhancement of long-lived excitonic correlations. In this case also the photoemission spectrum exhibits long-lived in-gap states related to these excitons.
The formation of excitons in the Mott gap due to the Coulomb interaction has been measured and discussed in the context of nonequilibrium optical experiments on 1D organic salts such as the Mott insulator ET-F$_2$TCNQ~\cite{mitrano2014}. 
However, such excitons have not been seen so far in pump-probe photoemission experiments. Our work suggests a promising new way of studying Mott excitons out-of-equilibrium in 1D and 2D materials.

Excitons may play a role in several recent experiments on photo-doped Mott insulators~\cite{novelli2014,terashige2019}. The mechanism underlying the exciton formation in this work is different from Ref.~\onlinecite{terashige2019}, which observed nearest-neighbor doublon-holon pairs in an antiferromagnetic background that are bound via the spin-spin interaction $J_\text{ex}$. In our system, $J_\text{ex}\approx \frac{4t_h^2}{U}=0.16$ is much smaller than the values of $V$ considered.  Material-specific modeling is required to assess the relevance of the two excitonic mechanisms for a given system. However, the generically large values of $V$ in realistic materials~\cite{miyake2008} suggest that doublon-holon pairs bound by the long-range Coulomb interaction and related spectral signatures should be taken into account in the analysis of photo-excited Mott insulators. 

In this work, we have focused on the dynamics of the photo-induced excitonic states in correlated systems due to non-local interaction. An important open question is how this excitonic dynamics is modified in the presence of photo-induced band-gap renormalizations~\cite{novelli2014,peli2017,golez2019dp,golez2019multi}. To address this issue, an EDMFT extension of the DCA formalism is required.

\begin{acknowledgments}
	This work was supported by ERC Consolidator Grant No.~724103 (NB, DG, PW), ERC starting grant No. 716648 (ME), and Swiss National Science Foundation Grant No.~200021\_165539 (NB, DG, PW). The calculations have been performed on the Beo04 cluster at the University of Fribourg. The Flatiron Institute is a division of the Simons Foundation. We thank T.~Tohyama, Y.~Murakami, and M. A.~Sentef for helpful discussions.
\end{acknowledgments}

\appendix

\section{Dynamical Cluster Approximation}

\subsection{Method}
\label{sec:Method}

\begin{figure}[!htb]
\centering
\includegraphics[width=0.99\columnwidth]{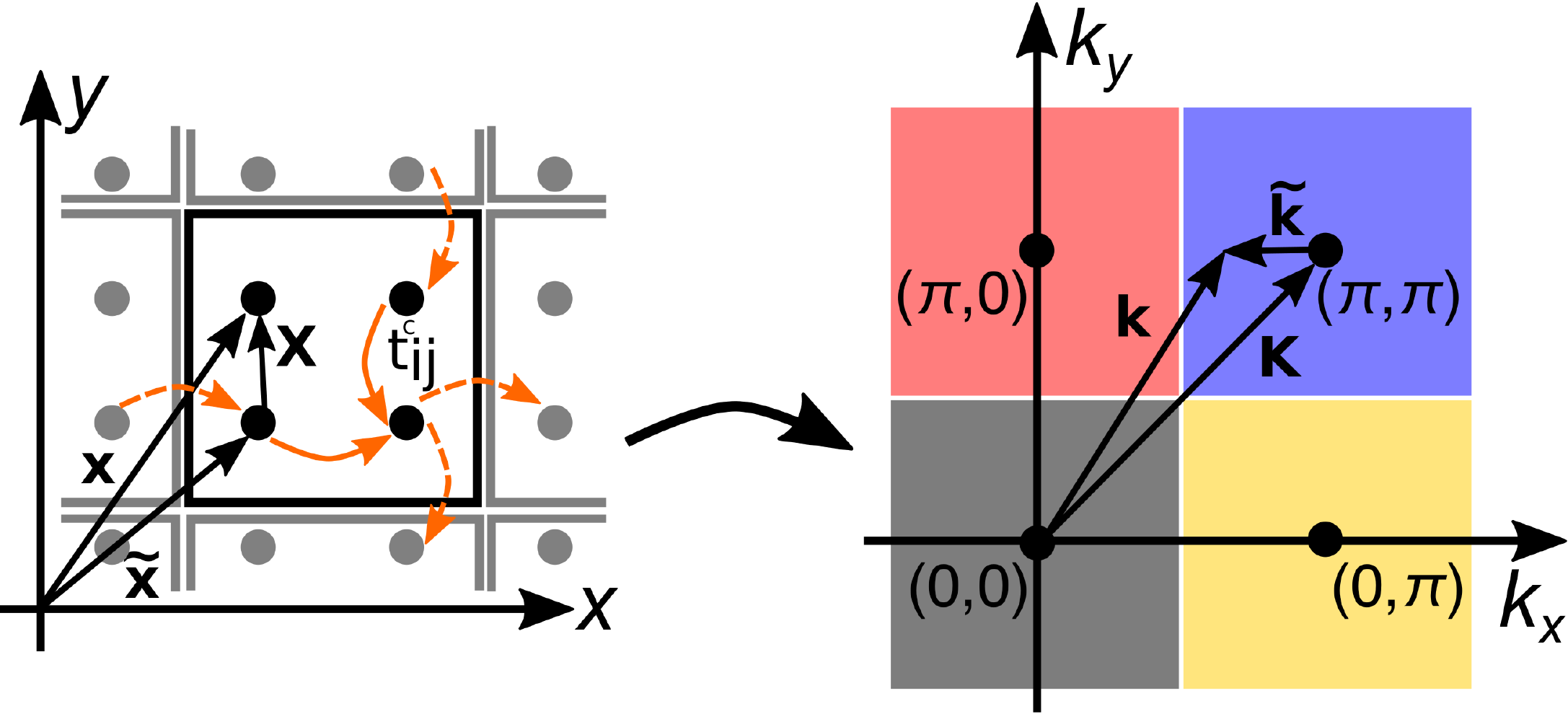}
\caption{(color online) Left panel: illustration of a $2\times2$ cluster in real space for the square lattice. Right panel: corresponding reciprocal space with $N_c=4$ patches.
}
\label{fig:Kpatch}
\end{figure}

Let us divide the 2D lattice in the real space into several clusters with $N_c$ being the number of sites within a cluster. We define the position of a site on the lattice by~\cite{maier2005}
\beq{
	\vx=\tilde{\vx}+\vX
}
with the superlattice vector $\tilde{\vx}$ and intra-cluster vector $\vX$. The corresponding vectors in the reciprocal space are $\vk$, $\tilde{\vk}$, $\vK$, respectively. The Brillouin zone is then divided into $N_c$ patches $P_\vK$ around the momentum vectors $\vK$. In our case we use a $2\times2$ cluster, which gives $N_c=4$ patches in the reciprocal space around $\vK=\left\{\left(0,0\right), \left(\pi,0\right), \left(0,\pi\right), \left(\pi,\pi\right)\right\}$, as shown in Fig.~\ref{fig:Kpatch}. In the DCA formalism the self-energy is assumed to be constant within each patch:
\beq{
	\Sigma_\vk(t,t')=\sum_{\vK} \Theta_{\vk,\vK}\Sigma_\vK(t,t'),
}
with $\Theta_{\vk,\vK}=1$ for $\vk$ in patch $\vK$ and zero otherwise.

In terms of the action $S$, the grand-canonical partition function can be written as
$\mathcal{Z}=\mathrm{Tr}[\mathcal{T_C} e^{S}]$ with $\mathcal{T_C}$ the contour-ordering operator
on the Kadanoff-Baym contour $\mathcal{C}$~\cite{aoki2014_rev}. 
The auxiliary cluster impurity problem for the extended Hubbard model Eq.~\eqref{eq:EHM} can be expressed
as a coherent-state path integral $\mathcal{Z}^c=\int   D[d^{*}, d]e^{S^c}$ with the action
\bsplit{
	S^c[d_i,d^*_i]=&-i\left\{\int_\mathcal{C} dt \sum_{\<i,j\>\sigma}[t^{c}_{i,j} d_{i\sigma}^*(t)d_{j\sigma}(t)+h.c.]\right.\\
	&\left. +\int_\mathcal{C} dt U\sum_{i}  n_{i\up}(t)n_{i\dn}(t) -\mu\int_C dt \sum_{i} n_{i}(t)\right.\\
&+\int_\mathcal{C} dt \sum_{{\<i,j\>, \sigma\sigma'}}  V_{i,j}^c d_{i\sigma}^*(t)d_{i\sigma}(t)d_{j\sigma'}^*(t)d_{j\sigma'}(t)\\
&\left.+\int_\mathcal{C} dt dt' \sum_{i,j,\sigma} d^*_{i\sigma}(t) \Delta_{i,j,\sigma}(t,t') d_{j\sigma}(t')\right\},
\label{eq:dca_realspace}
}
with renormalized hopping matrix elements $t_{i,j}^c$ and inter-site interactions $V_{i,j}^c$, as defined below.  $\Delta_{i,j,\sigma}$ is the hybridization function, which is calculated self-consistently and in real space contains off-diagonal elements.

Since the cluster is periodized in DCA, it is convenient to perform a Fourier transformation
\bsplit{
\label{eq:FTdca}
	d_{\vK\sigma}^*(t)&=\frac{1}{\sqrt{N_c}}\sum_{i}e^{-i\vX_i\cdot\vK}d_{i\sigma}^*(t),\\
	d_{\vK\sigma}(t)&=\frac{1}{\sqrt{N_c}}\sum_{i}e^{i\vX_i\cdot\vK}d_{i\sigma}(t),\\
	\Delta_{\vK\sigma}(t,t')&=\frac{1}{N_c}\sum_{i,j}e^{i(\vX_i-\vX_j)\cdot\vK}\Delta_{i,j,\sigma}(t,t'),
}
and to express the action in reciprocal ($\vK$) space, where $\Delta_{\vK\sigma}(t,t')$ is diagonal in $\vK$.
The Green's functions and self-energies are also diagonal in $\vK$ space.

The cluster action written in momentum space becomes
\bsplit{
	S^c[d_{\vK\sigma},d^*_{\vK\sigma}]=&-i\int_\mathcal{C}dt H^c(t)\\
	& -i\int_\mathcal{C}dtdt'\sum_{\vK \sigma}d^*_{\vK\sigma}(t)\Delta_{\vK\sigma}(t,t')d_{\vK\sigma}(t'),
}
with
\bsplit{
	H^c(t)=&\sum_{\vK,\sigma}\bar{\epsilon}_\vK(t) d_{\vK\sigma}^* d_{\vK\sigma}-\mu^c\sum_{\vK,\sigma} d_{\vK\sigma}^*d_{\vK\sigma} \\
	&+\frac{U}{N_c}\sum_{\vK,\vK',\vQ}d^*_{\vK\up}d^*_{\vK'\dn}d_{\vK'-\vQ\dn}d_{\vK+\vQ\up}\\
	&+\frac{1}{N_c}\sum_{\substack{\vK,\vK',\vQ\\ \sigma, \sigma'}}\bar V_\vQ d^*_{\vK\sigma}d^*_{\vK'\sigma'}d_{\vK'+\vQ\sigma'}d_{\vK-\vQ\sigma}.
}
Here, $\bar{\epsilon}_{\vK}=\frac{N_c}{N}\sum_{\vk}\Theta_{\vk,\vK}\epsilon_\vk$
denotes the coarse-grained dispersion and $\bar V_{\vQ} = \frac{N_c}{N}\sum_{\vq}\Theta_{\vq,\vQ} V_\vq$
the coarse-grained nearest neighbor interaction. In our case of a 2D square lattice with nearest-neighbor hoppings and interactions, we have $\epsilon_\vk (t)=-2t_h(t)\left(\cos k_x + \cos k_y\right)$, $V_\vq=V(\cos q_x+\cos q_y)$ and
$\bar \epsilon_\vK (t)=-2 \frac{2}{\pi} t_h(t) \left(\cos \vK_x + \cos \vK_y\right)$, $\bar V_\vQ=\frac{2}{\pi}V(\cos \vQ_x+\cos \vQ_y)$.
The backward Fourier transforms of $\bar{\epsilon}_{\vK}$ and $\bar V_{\vQ}$ define the renormalized parameters $t^c_{i,j}=\frac{2}{\pi}t_h$ and $V^c_{i,j}\equiv V^c=\frac{2}{\pi}V$ in Eq.~(\ref{eq:dca_realspace}) for nearest neighbor sites $i$ and $j$.
The half-filling condition for the cluster is $\mu^c=U/2+4V^c$.

\begin{figure}[!t]
\centering
	\includegraphics[width=1.0\columnwidth]{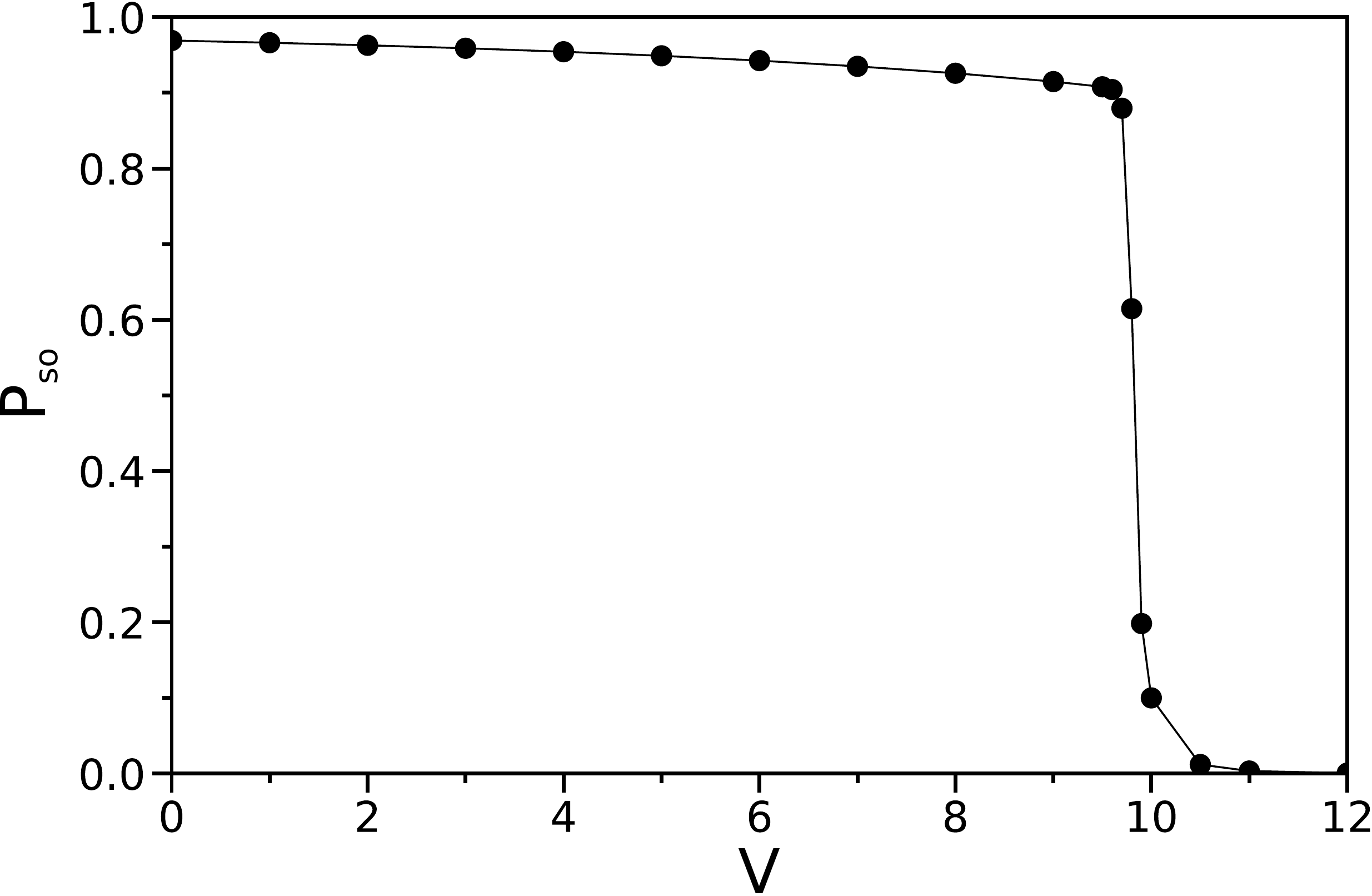}
\caption{(color online) Correlation function $\Pso$, which measures singly occupied sites on the cluster, versus  nearest-neighbor interaction $V$ for $U=25$ and $T=0.1$.
}
\label{fig:appPso}
\end{figure}

The solution of the cluster impurity model yields
the cluster Green's function $G^c_{\vK}(t,t') = -i \langle d_\vK(t)d_\vK^\+(t')\rangle$.  It implicitly defines the cluster self-energy $\Sigma_\vK$ via the Dyson equation
\begin{equation}
\label{eq:DySigma}
\begin{split}
	[G^c_\vK]^{-1}(t,t')=& [i\partial t +\mu^c - \bar{\epsilon}_\vK(t)]\delta(t,t') \\
	&- \Delta_{\vK\sigma} (t,t') - \Sigma_\vK (t,t'),
\end{split}
\end{equation}
which may then be used to calculate the lattice Green's function $\tilde{G}_{\vk}(t,t') = -i \langle c_{\vk}(t)c_{\vk}^\+(t')\rangle$ using the lattice Dyson equation
\beq{
	[\tilde{G}_{\vk}]^{-1}(t,t')=[i\partial t+\mu^c -\epsilon_\vk(t)]\delta(t,t') - \Sigma_\vk(t,t').
}
In the calculations we use a square grid of $20\times20$ $\vk$ points.
The fermionic self-consistency is closed by coarse-graining of the lattice Green's function
\beq{
	G^c_\vK(t,t')=\frac{N_c}{N}\sum_{\vk\in P_\vK}\tilde{G}_\vk(t,t').
}
We do not include a bosonic self-consistency loop in our formalism, i.e., the on-site interaction $U$ remains unscreened within this approximation.

Since we are considering Mott insulating systems, the simulations employ an impurity solver based on the non-crossing approximation~\cite{grewe1981, coleman1984, eckstein2010} (NCA). In combination with this solver, it is more convenient to work with the quantity $\Lambda=1/(i\partial_t-\Sigma)$ instead of $\Sigma$ and to close the fermionic self-consistency loop in a manner analogous to what has been described for the single-orbital case in Refs.~\onlinecite{eckstein2011,aoki2014_rev}.

\subsection{Calculation of the correlation functions using DCA}
\label{sSec:CorrFct}

Since the DCA self-consistency loop is formulated in momentum space, it is convenient to measure also the double occupation and the correlation functions in $\vK$ space.
For instance, let us consider the double occupancy function:
\beq{
	D(t) = \frac{1}{N_c}\sum_{i=1}^{N_c} \<\hn_{i\up} \hn_{i\dn}\>(t)
	= \frac{1}{N_c}\sum_{i} \<d_{i\up}^\+d_{i\up}d_{i\dn}^\+d_{i\dn}\>(t).
}
Using the Fourier transformation~\eqref{eq:FTdca} and the relation $\frac{1}{N_c}\sum_i  e^{-i(\vK_1-\vK_2+\vK_3-\vK_4)\cdot \vR_i}=\delta_{\vK_4, \vK_1-\vK_2+\vK_3}$
the double occupancy function $D(t)$ can be expressed as
\beq{
	D(t) = \frac{1}{N_c^2}\sum_{\vK \vK' \vQ} \<d_{\vK\up}^\+d_{\vK'\dn}^\+d_{\vK'-\vQ \dn}d_{\vK+\vQ \up}\>(t).
}

\begin{figure}[!b]
\centering
	\includegraphics[width=1.0\columnwidth]{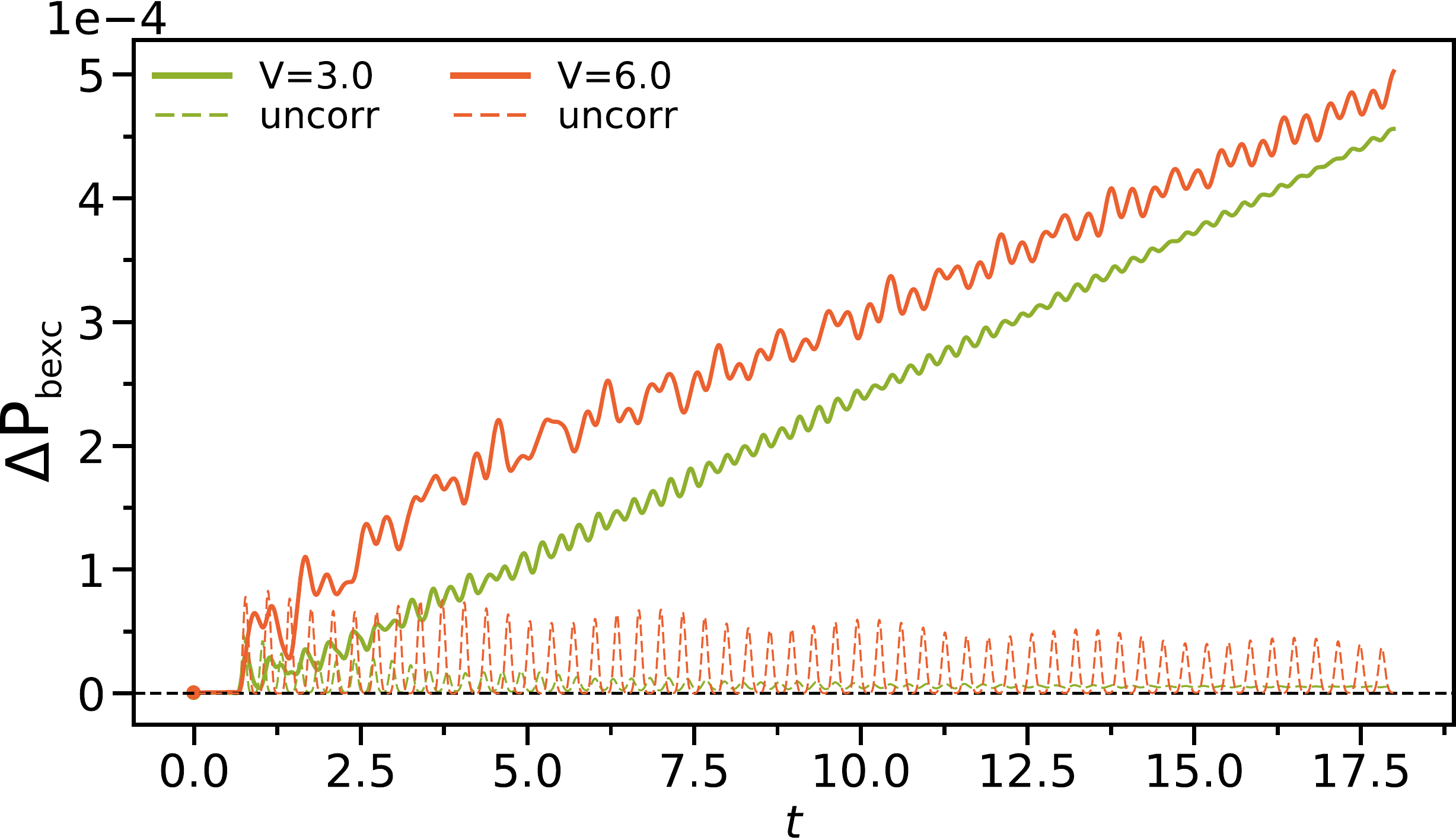}
\caption{(color online) Temporal evolution of the bi-excitonic correlation function for $V=3$ (green solid line) and for $V=6$ (red solid line) in a system with $U=25$. The parameters of the excitation pulse are the same as in Fig.~\ref{fig:tCorrV03} and in Fig.~\ref{fig:tCorrV06}. The corresponding probability for uncorrelated pairs of doublons and holons on the cluster (see Eq.~\eqref{eq:uncorrPbexc}) is illustrated by the dashed lines.
}
\label{fig:tPbexc}
\end{figure}

\section{Additional equilibrium properties}

\begin{figure*}[!t]
\centering
	\includegraphics[width=1.0\textwidth]{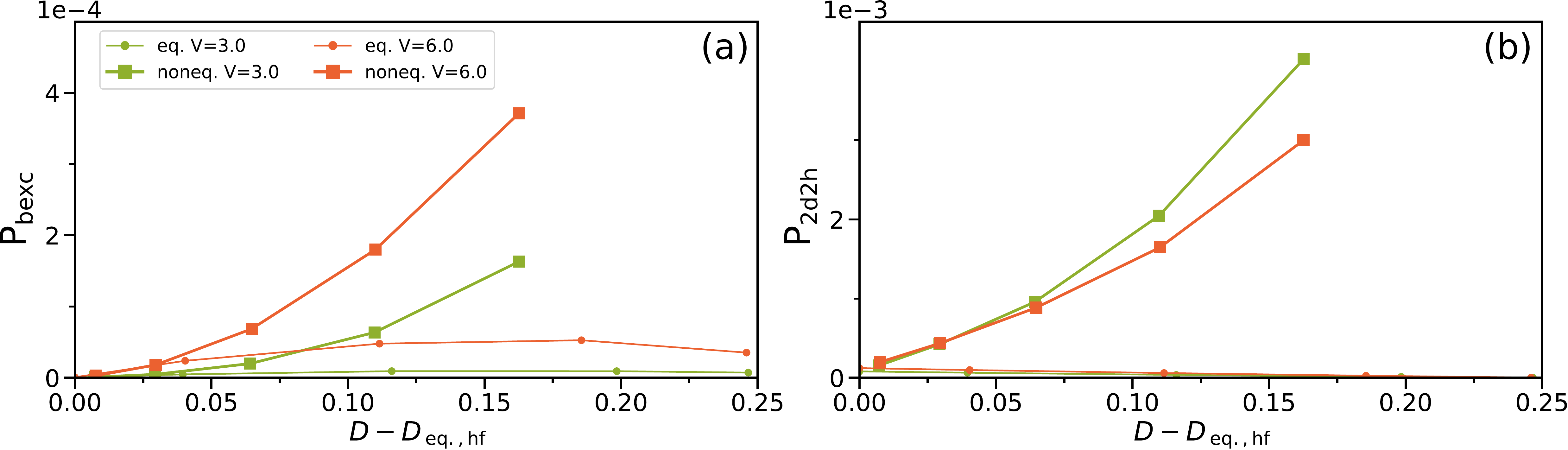}
\caption{(color online) Comparison of photo-doped and chemically doped systems: (a) $\Pbexc$, and (b) $\P2d2h$ versus double occupancy measured relative to the equilibrium half-filled value. The calculations are done for $U=25$ at temperature $T=0.1$. Different colors correspond to different values of the next-neighbor interaction $V$.  Dots represent equilibrium results, whereas squares indicate the results of the nonequilibrium calculations.
}
\label{fig:Pnoneq_appendix}
\end{figure*}

\subsection{Correlation function for singly occupied sites}
\label{sec:appEq}

We define the correlation function $\Pso$, which measures the probability for four singly occupied sites on the cluster,
\begin{equation}
\label{eq:Pso}
	\begin{split}
		\Pso(t) =& \frac{1}{4}\sum_{i} \<(\hn_i-2\hD_i) (\hn_{i+1}-2\hD_{i+1})\times\right.\\
	&\times \left. (\hn_{i+2}-2\hD_{i+2})(\hn_{i+3}-2\hD_{i+3}) \>(t).
	\end{split}
\end{equation}
In Fig.~\ref{fig:appPso}, we plot $\Pso$ for $U=25$ as a function of the nearest neighbor interaction strength $V$.
While for $V\lesssim9.7$ this correlation function takes values close to 1, in agreement with a dominant plaquette singlet state~\cite{gull2008}, for larger nearest-neighbor interactions in the ``CO" regime, $\Pso$ gets strongly suppressed.

\subsection{Exciton energy}
\label{Ssec:appEexc}

Let us focus for simplicity on an isolated periodized dimer in the half-filled case. Further, we assume a system in the AFM regime and the half-filling condition $\mu=U/2+2V^c$. This system exhibits 4 eigenstates, which in the large-$U$ limit can be approximately written as 
\begin{equation}
	\begin{split}
		\ket{\text{GS}}\equiv\ket{0}&=\tfrac{1}{\sqrt{2}}(\ket{\up\quad\dn}-\ket{\dn\quad\up}),\\
		\ket{1}&=\tfrac{1}{\sqrt{2}}(\ket{\up\quad\dn}+\ket{\dn\quad\up}),\\
		\ket{2}&=\tfrac{1}{\sqrt{2}}(\ket{\up\dn\, \mathrm{o}}-\ket{\mathrm{o}\, \up\dn}),\\
		\ket{3}&=\tfrac{1}{\sqrt{2}}(\ket{\up\dn\, \mathrm{o}}+\ket{\mathrm{o}\, \up\dn}),
	\end{split}
\end{equation}
where the configuration $\ket{\up\quad\dn}$ represents two electrons with spins $\up$ and $\dn$ sitting on the two different sites of the dimer and the configuration $\ket{\up\dn\, \mathrm{o}}$ corresponds to a doubly occupied ($\up\dn$) and an empty ($\mathrm{o}$) site.
The corresponding eigenenergies are 
\begin{equation}
	\begin{split}
		E_{0}&=-U/2-3V^c-\sqrt{(U/2-V^c)^2+16t_h^2},\\
		E_{1}&=-U-2V^c,\\
		E_{2}&=-4V^c,\\
		E_{3}&=-U/2-3V^c+\sqrt{(U/2-V^c)^2+16t_h^2}.
	\end{split}
\end{equation}
Hence, the energy difference between the excitonic state $\ket{2}$ and the ground state $\ket{\text{GS}}$ is:
\begin{equation}
	E_\mathrm{exc}=E_2-E_0=\frac{U}{2}-V^c+\sqrt{\left(\frac{U}{2}-V^c\right)^2+16t_h^2}.
\end{equation}
Since in the AFM phase $V^c<U/2$, we can express this as
\begin{equation}
	E_\mathrm{exc}=\left(\frac{U}{2}-V^c\right)\left(1+\sqrt{1+\frac{16t_h^2}{\left(\frac{U}{2}-V^c\right)^2}}\right).
\end{equation}
Further simplification for $U/2-V^c>4t$ using the Taylor expansion leads to the final result
\begin{equation}
	E_\mathrm{exc}\approx U-2V^c+\frac{16t_h^2}{U-2V^c}.
\end{equation}

\section{Additional time-dependent correlation functions}

\label{sec:apptCorr}

In Fig.~\ref{fig:tPbexc}, we plot the temporal evolution of the bi-excitonic correlation function $\Pbexc(t)$ after a hopping modulation for $U=25$, $V=3$ (green line) and $V=6$ (red line). The correlation function  is measured with respect to its equilibrium value. We note that the equilibrium value of $\Pbexc$ is tiny (${\cal O}(10^{-5})$, not shown) for both nearest-neighbor interaction strengths. The photo-excitation (which lasts up to $t\approx 1.2$) leads to a systematic enhancement in $\Delta\Pbexc$. However, the photo-induced values of $\Pbexc$ remain tiny.

Next, we calculate the probability for uncorrelated pairs of doublons and holons on the cluster, which we calculate in analogy to Eqs.~\eqref{eq:uncorrPexc} and~\eqref{eq:uncorrPexcNNN} as
\begin{equation}
\label{eq:uncorrPbexc}
	\Pbexc^\text{uncorr}(t) = 2D^4(t).
\end{equation}
The comparison of $\Delta \Pbexc^\text{uncorr}(t)$ (dashed lines in Fig.~\ref{fig:tPbexc}) with $\Delta\Pbexc$ indicates pulse-induced bound pairs of doublons and holons on the diagonals of the cluster.

To analyse the dependence of the bi-excitonic correlation function on the chemical doping (dots) and on the photo-doping (squares), we plot in Fig.~\ref{fig:Pnoneq_appendix}(a) $\Pbexc$ as a function of the change in the double occupancy relative to the equilibrium half-filled value. The parameters of the system and of the excitation process are the same as used in Fig.~\ref{fig:Pnoneq}. In addition, we also define a correlation function, which measures the probability of having two neighboring doublons and two neighboring holons on the cluster:
\begin{equation}
\label{eq:P2d2h}
	\P2d2h(t) = \sum_{i} \<\hD_i \hD_{i+1} \hh_{i+2} \hh_{i+3}\>(t).
\end{equation}
The corresponding dependence on the chemical doping and on the photo-doping is shown in Fig.~\ref{fig:Pnoneq_appendix}(b). As one can see, while the chemical doping suppresses $\P2d2h$ and only slightly enhances $\Pbexc$, the photo-doping leads to a clear enhancement of both values. Interestingly, $\P2d2h$ is an order of magnitude larger than $\Pbexc$, although still small in comparison to $\Pexc$. We also note that the correlations of the paired doublons and holons on the cluster get suppressed with increasing nearest neighbor interaction $V$, in contrast to $\Pbexc$.

\bibliography{bibtex/Polarons,bibtex/tdmft,bibtex/Books,bibtex/tools}
\vspace{0cm}
\mbox{}

\end{document}